\documentclass[fleqn,usenatbib]{mnras}

\usepackage[T1]{fontenc}
\usepackage{booktabs}
\DeclareRobustCommand{\VAN}[3]{#2}
\let\VANthebibliography\thebibliography
\def\thebibliography{\DeclareRobustCommand{\VAN}[3]{##3}\VANthebibliography}

\usepackage{graphicx}	
\usepackage{amsmath}	
\usepackage{amssymb}	
\usepackage[table]{xcolor}
\usepackage{multirow}
\usepackage{newtxtext,newtxmath}

\newcommand{\thickhline}{%
    \noalign {\ifnum 0=`}\fi \hrule height 1pt
    \futurelet \reserved@a \@xhline
}

\defcitealias{CC02}{CC02}
\defcitealias{CM02}{CM02}

\title[kSZ features from cluster rotation]{Galaxy cluster rotation revealed in the MACSIS simulations with the kinetic Sunyaev-Zeldovich effect}

\author[E. Altamura et al.]{
Edoardo Altamura,$^{1}$\thanks{E-mail: \href{mailto:edoardo.altamura@manchester.ac.uk}{edoardo.altamura@manchester.ac.uk}}
Scott T. Kay,$^{1}$\thanks{E-mail: \href{mailto:scott.kay@manchester.ac.uk}{scott.kay@manchester.ac.uk}}
Jens Chluba$^{1}$
and Imogen Towler$^{1}$
\\
$^{1}$Jodrell Bank Centre for Astrophysics, Department of Physics and Astronomy, The University of Manchester, Oxford Road, Manchester M13 9PL, UK\\
}

\date{Accepted XXX. Received YYY; in original form ZZZ}

\pubyear{2023}

\begin{document}
\label{firstpage}
\pagerange{\pageref{firstpage}--\pageref{lastpage}}
\maketitle

\begin{abstract}
The kinetic Sunyaev-Zeldovich (kSZ) effect has now become a clear target for ongoing and future studies of the cosmic microwave background (CMB) and cosmology. Aside from the bulk cluster motion, internal motions also lead to a kSZ signal. In this work, we study the rotational kSZ effect caused by coherent large-scale motions of the cluster medium using cluster hydrodynamic cosmological simulations. To utilise the rotational kSZ as a cosmological probe, simulations offer some of the most comprehensive data sets that can inform the modelling of this signal. In this work, we use the MACSIS data set to investigate the rotational kSZ effect in massive clusters specifically. Based on these models, we test stacking approaches and estimate the amplitude of the combined signal with varying mass, dynamical state, redshift and map-alignment geometry. We find that the dark matter, galaxy and gas spins are generally misaligned, an effect that can cause a sub-optimal estimation of the rotational kSZ effect when based on galaxy {motions}. Furthermore, we provide halo-spin-mass scaling relations that can be used to build a statistical model of the rotational kSZ. The rotational kSZ contribution, which is largest in massive unrelaxed clusters ($\gtrsim$100 $\mu$K), could be relevant to studies of higher-order CMB temperature signals, such as the moving lens effect. The limited mass range of the MACSIS sample strongly motivates an extended investigation of the rotational kSZ effect in large-volume simulations to refine the modelling, particularly towards lower mass and higher redshift, and provide forecasts for upcoming cosmological CMB experiments (e.g. Simons Observatory, SKA-2) and X-ray observations (e.g. \textit{Athena}/X-IFU).
\end{abstract}

\begin{keywords}
Hydrodynamics -- 
methods: miscellaneous --
methods: statistical --
galaxies: clusters: intracluster medium --
galaxies: kinematics and dynamics --
cosmology: observations
\end{keywords}



\section{Introduction}
The kinetic Sunyaev-Zeldovich (kSZ) effect is {related to} a late-{time} Doppler boost of the cosmic microwave background (CMB) which manifests itself when ionised gas moves with a non-zero velocity in the CMB rest-frame \citep{ksz_sunyaev_1980}. The kSZ has been investigated theoretically and observationally \citep[e.g.][]{2020PhRvL.125k1301C, 2022MNRAS.510.5916C} {on both cosmological and astrophysical levels.} In the cosmological context, {the kSZ contribution from astrophysical CMB foregrounds can be used to track the prokected velocity field in large-scale structures \citep[e.g.][]{2023JCAP...03..039B} or} subtracted to retrieve the signal from CMB anisotropies \citep[e.g.][]{2020A&A...641A...4P}. In the astrophysical context, the kSZ signal is a direct and unique probe for the dynamical state of the intra-cluster medium (ICM) in galaxy clusters \citep{mroczkowski2019_review}. {In addition to} direct measurements of the kSZ effect due to the {pair-wise momentum of thousands of} clusters\footnote{{In \cite{2012PhRvL.109d1101H}, the objects were selected based on 27291 luminous galaxies from the Baryon Oscillation Spectroscopic Survey Data Release 9 \citep[BOSS-DR9,][]{2011ApJ...728..126W}. Atacama Cosmology Telescope \citep[ACT,][]{2011ApJS..194...41S} observations provided the microwave data to measure the temperature distortions.}} \citep{2012PhRvL.109d1101H}, more sensitive instruments and sophisticated post-processing pipelines have enabled the detection of peculiar motion of individual substructures within {an individual} cluster \citep{2017A&A...598A.115A}.

While galaxy clusters in quasi-hydrostatic equilibrium are mostly pressure-supported, they gain angular momentum from the surrounding matter during their gravitational collapse and maintain a residual rotational support, typically accounting for {$\simeq 5\%$} of their total kinetic energy \citep{1995MNRAS.272..570S, 1996MNRAS.281..716C}. The presence of ordered motions in the ICM affects the assumption of hydrostatic equilibrium, often adopted to estimate cluster masses \citep[see e.g.][]{2005ApJ...628..655V} and the ability to quantify this discrepancy may improve the current estimates on hydrostatic mass bias and the angular momentum distribution of large assemblies of galaxies. Using the galaxy cluster Abell 2107 as a case-study, \cite{2005MNRAS.359.1491K} illustrated that accounting for bulk rotation can lead to a few 10s percentile differences in the mass estimates and can help reconstruct the recent dynamical history of the system \citep{2019MNRAS.485.3909L} and the connection to the surrounding large-scale structures \citep{2018ApJ...869..124S}. 

Using very similar assumptions, \citet{CC02} [henceforth, \citetalias{CC02}] and \citet{Chluba2001Diploma, CM02} [henceforth, \citetalias{CM02}] estimated the additional kSZ signal deriving from ordered cluster rotation, which combines with that from the cluster's bulk peculiar velocity. This effect, known as rotational kSZ [or rkSZ, not to be confused with the \textit{relativistic} SZ, see e.g. \citet{Sazonov1998, Challinor1998, Itoh98, Chluba2012SZpack, 2020MNRAS.493.3274L}], was initially estimated to produce a temperature variation $\Delta T_{\rm rkSZ}$ over the CMB ranging from $\simeq 3.5\, \mu$K for a relaxed cluster to $\simeq 146\, \mu$K for a recent merger \citepalias{CM02}, assuming a halo $\beta$-model from \cite{1976A&A....49..137C} and solid body rotation. The rkSZ signal has not yet been observed in individual clusters due to its remarkably small amplitude and its dependence on the orientation of the rotation axis relative to the line of sight (LoS).

The ICM gas moving towards the observed produces a temperature increment over the CMB, while gas moving in the opposite direction leads to a  temperature {decrement}. In the presence of cluster rotation, these patterns are adjacent and produce a dipole-like signature, which in the general case is superimposed to the monopole-like signal due to the cluster's bulk motion along the LoS. {Such} dipolar patterns in the CMB temperature map can also arise from the gravitational moving-lens effect {\citep{1986Natur.324..349G, 2007MNRAS.380.1023S, 2021PhRvD.104h3529H, 2021PhRvD.103d3536H}}: the CMB photons, deflected by the deep gravitational potential well of clusters, cause the anisotropies to be re-mapped and imprint an additional dipole-like feature in the temperature distribution. The dipole-like pattern from weak lensing has a temperature and angular scale comparable to that from the rkSZ effect {(see \citealt{2000ApJ...538...57S}, Section 4.3.1 of \citealt{2015ApJ...806..247B}, and \citealt{2019PhRvL.123r1301R})}, making it challenging to distinguish the two effects.

Recently, attempts to isolate the the rkSZ signal from the hot circumgalactic medium of 2000 galaxies were performed by \citet{2020PhRvD.101h3016Z}. {Slightly earlier,} \citet{2019JCAP...06..001B} presented a similar analysis of {\textit{Planck} data using 13 galaxy clusters from the SDSS-DR10 \citep{2014ApJS..211...17A} showing indications of bulk rotation \citep{2017MNRAS.465.2616M}}. In both works, the authors state the importance of \textit{aligning} and \textit{stacking} the kSZ maps from multiple objects to retrieve the rotational signal with sufficient signal-to-noise ratio. Because the rkSZ signal produces a dipolar pattern in the observed $\Delta T$ field, the maps must be {oriented} such that the {projected} rotation axis of the objects in the sample is aligned to maximise the rotational signal. The scale of the maps is then normalised to the objects' self-similar scale radii and the results are finally stacked. 

Synthetic galaxy clusters produced in hydrodynamic simulations offer unique test-cases for predicting the kSZ signal from bulk motion and rotation of the ICM. Crucially, simulations model the formation of clusters from cosmological accretion and therefore can capture the angular momentum transfer during gravitational collapse, mergers and substructures, all of which are not included in the analytic models used by \citetalias{CM02}. Using six clusters selected from the MUSIC simulations \citep{2013MNRAS.429..323S}, \citet{2018MNRAS.479.4028B} found that the rkSZ signal can account for up to 23\% of the kSZ component purely from bulk motion. They extend the study by showing the rkSZ signal variation at different orientations of the rotation axis relative to the LoS. Using a 6-parameter \citet{2006ApJ...640..691V} model fit to the electron number density profile of each halo and a parametric tangential velocity profile from \cite{2017MNRAS.465.2584B}, they could recover the tangential scale-velocity and bulk velocity by fitting the analytic model to the synthetic kSZ maps. Since the work by \citet{2018MNRAS.479.4028B}, \citet{2021MNRAS.504.4568M} have produced rkSZ maps to improve the accuracy of halo spin bias estimates using $5\times10^4$ halos with virial mass between $1.48 \times 10^{11}$ M$_\odot$ and $4.68 \times 10^{14}$ M$_\odot$ drawn from the IllustrisTNG simulation \citep{2018MNRAS.473.4077P}.

In this paper, we focus our discussion on the rotation of clusters, with particular reference to the work by \citet{2019JCAP...06..001B}. To obtain the rotation axis of the galaxy clusters in their sample, \citet{2019JCAP...06..001B} matched the {\textit{Planck} SZ maps with SDSS-DR10 galaxies in clusters showing evidence of coherent rotation based on the {LoS} velocity of the sources \citep{2017MNRAS.465.2616M}}. {Under the assumption that the galaxies and the ICM rotate about} the same axis, they appropriately oriented, scaled and then stacked the \textit{Planck} SZ maps to maximise the amplitude of the dipole-like signature of the kSZ effect from cluster rotation. Using the rkSZ analytic model from \citetalias{CM02}, they estimated the model parameters using the maximum likelihood estimation method, {yielding $\simeq 2\, \sigma$ evidence for the presence of the effect}.

In a $\Lambda$-cold dark matter (CDM) cosmology, the collapse of structures and the shape of the gravitational potential at low redshift is dominated by dark matter, with the baryonic matter following the same evolution. This concept implies that the different components of galaxy clusters (ICM gas, dark matter and stars in galaxies) are expected to co-rotate and to have their total angular momenta aligned. Early simulations found that the spin of dark matter halos is usually well-aligned with that of the central galaxy and the ICM, except for a non-negligible fraction of the population of objects showing misalignment \citep{2002ApJ...576...21V, 2010MNRAS.404.1137B}. The same works found that the angular momentum orientation of matter in the inner and outer halo is often vastly different, suggesting the importance in choosing an appropriate aperture when defining the spin of a galaxy cluster. More recently, these results have been corroborated using the Illustris simulation, which produced a surprisingly large ($30-50^{\circ}$ considering particles within the virial radius) median misalignment between gas and galaxies in cluster-sized objects, explained by the old-type stars being subject to the gravitational potential of the dark matter field, while being relatively unaffected by the gas hydrodynamics in the ICM in the late-time halo assembly \citep[see section 5.5 of][]{2017MNRAS.466.1625Z}. Although this estimate for the gas-stars misalignment includes \textit{all stars} within the virial radius instead of just the satellite galaxies as in the set-up used by \citet{2019JCAP...06..001B}, the result from the Illustris simulation suggests that the co-rotation of stars and gas, critical for recovering the elusive rkSZ signal, requires further inspection.

In this work, we aim to investigate the assumption of co-rotating galaxies and ICM gas used by \citet{2019JCAP...06..001B}, modelling the rkSZ signal of massive clusters in different alignment and stacking scenarios. The MACSIS simulations \citep{macsis_barnes_2017} provide an excellent suite of synthetic galaxy clusters simulated with the the \textsc{Gadget-3} code and the BAHAMAS sub-grid physics model. The MACSIS clusters were selected to have FoF mass above $10^{15}$ M$_\odot$, which extends the sample in \citet{2017MNRAS.466.1625Z} and \citet{2018MNRAS.479.4028B} to higher masses by one order of magnitude. Crucially, the amplitude of the SZ effects is larger in massive halos, meaning that observational surveys are most likely to detect rotational features in the SZ sky by selecting massive and merger-prone MACSIS-like clusters \citep{CM02, 2003AstL...29..783S}.

This work is organised as follows. Section \ref{sec:datasets} introduces the key features of the MACSIS cluster sample and  the theoretical framework for the rkSZ effect; in Section \ref{sec:cluster-properties}, we examine the alignment of the angular momenta of cluster components; in Section \ref{sec:image-processing}, we reproduce the rkSZ map-stacking method used in \citet{2019JCAP...06..001B} and in Section \ref{sec:results} explore the rotational signal with different selection criteria. We then fit a analytic model to the profiles as discussed in Section~\ref{sec:results:fitting} and, {starting from the prescription of \citetalias{CC02},} in Section~\ref{sec:power-spectrum} we illustrate how {our} results can be used to predict the contribution of the cluster rotation to the kSZ power spectrum. Finally, in Section \ref{sec:summary}, we discuss prospects for future models and observations of the kSZ effect from cluster rotation.

Throughout this work, we adopt the cosmology used in MACSIS \citep{macsis_barnes_2017}, with parameters: $\Omega_{\rm b} = 0.04825$, $\Omega_{\rm m} = 0.307$, $\Omega_\Lambda = 0.693$, $h \equiv H_0/(100 ~{\rm km~s^{-1}Mpc^{-1}}) = 0.6777$, $\sigma_8 = 0.8288$, $n_s = 0.9611$ and $Y = 0.248$ \citep{2014A&A...571A...1P}.

\section{Overview of the MACSIS simulations}
\label{sec:datasets}

\begin{figure*}
	\includegraphics[width=2\columnwidth]{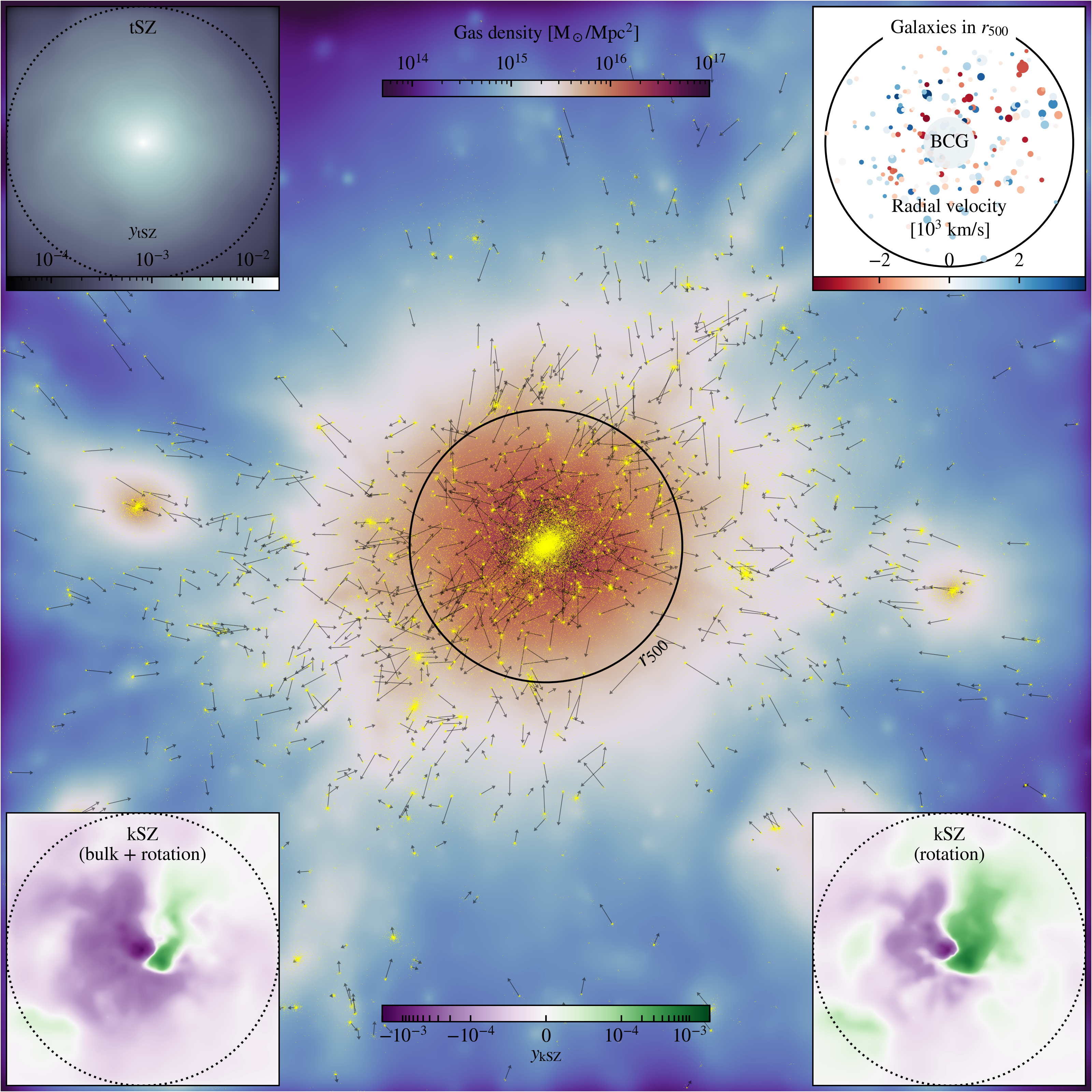}
    \caption{Illustration of the hot gas and stars in the MACSIS 0 cluster environment at $z=0$. The central image shows the projected gas density in the background (colour-coded), with superimposed the star particles marked in yellow. For the substructures with stellar mass above $10^{10}$ M$_\odot$, we also show the projected velocity vectors as black arrows. The spatial extent of the central map is 8 $r_{500}$, and the $r_{500}=2.39$ Mpc circle is drawn to guide the eye. The cluster is rotated such that the angular momentum of the hot gas in the ICM points vertically upwards in the plane of the page. The maps in the insets all have an extent of 2 $r_{500}$; the dotted circles in the three SZ maps indicate $r_{500}$. In the top-left, we show a map of the tSZ Compton-$y$ parameter; in the bottom-left, we show the kSZ Compton-$y$ parameter for the hot gas in the rest frame of the CMB; in the bottom-right is the same kSZ map, but without the cluster's bulk motion, as indicated by the label. The kSZ (bulk + rotation) and rotation-only amplitudes are comparable, however, we note that a large component of the bulk velocity of the cluster is oriented tangentially and, therefore, it does not contribute to the kSZ signal {at order $\simeq v/c$}. Finally, the top-right plot shows the position of the galaxies inside $r_{500}$ (3D, not projected), with markers colour-coded based on the {LoS} velocity and with size proportional to the logarithm of their stellar mass. The BCG is indicated in the centre of the plot, as well as the $r_{500}$ radius.}
    \label{fig:cluster_display}
\end{figure*}

\begin{figure*}
	\includegraphics[width=2\columnwidth]{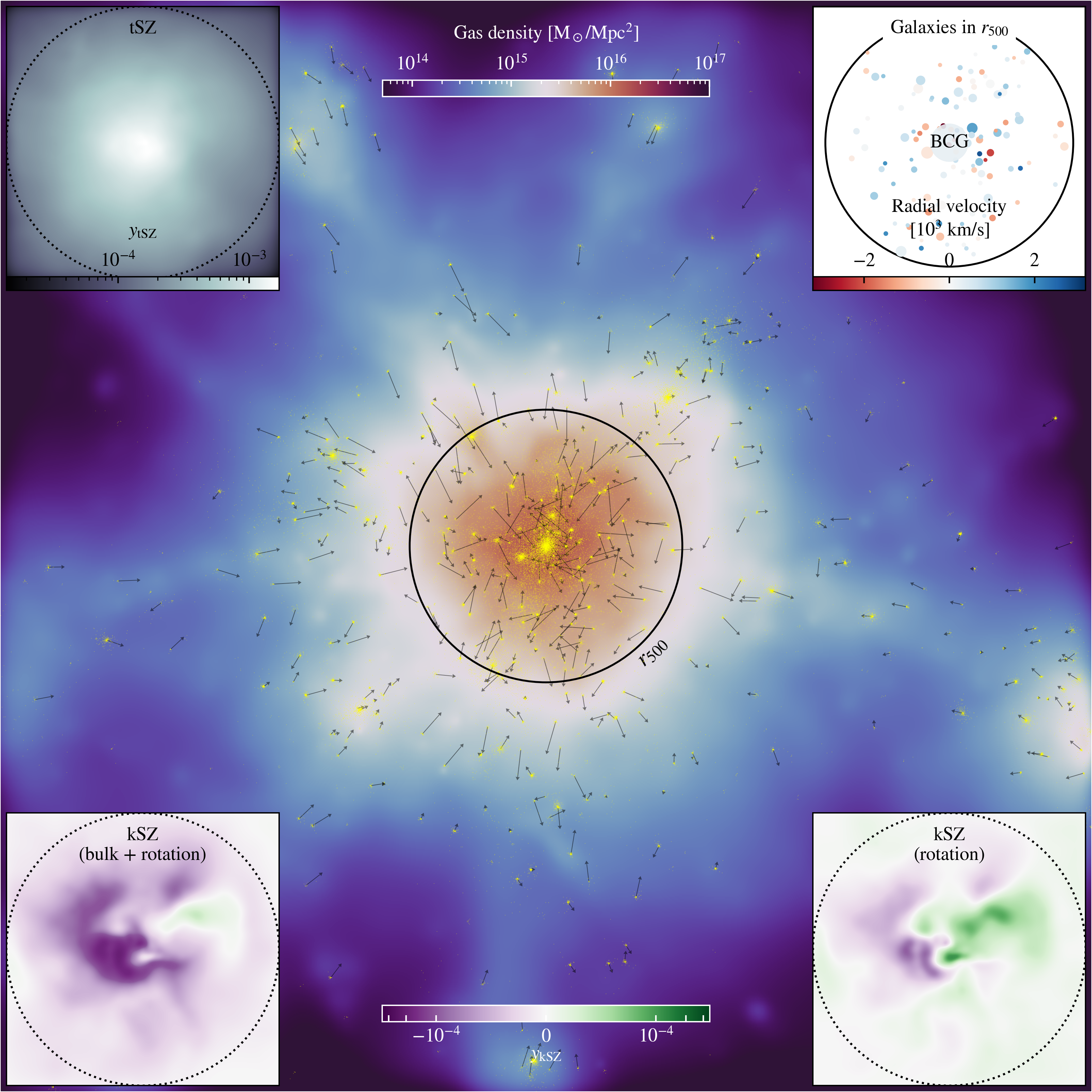}
    \caption{As in Fig. \ref{fig:cluster_display}, but showing the MACSIS 100 cluster at $z=0$. This smaller system, with $M_{500}=1.77 \times 10^{15}$ M$_\odot$ and $r_{500}=1.88$ Mpc, shows a lower number of galaxies with stellar mass above $10^{10}$ M$_\odot$. Kinematically, we report a significant bulk motion (bottom left) along the {LoS} and a residual dipolar kSZ pattern due to bulk rotation (bottom right). {The dipolar rotational signature is less pronounced due to the presence of moving substructures.}}
    \label{fig:cluster_display_100}
\end{figure*}

\begin{figure*}
	\includegraphics[width=2\columnwidth]{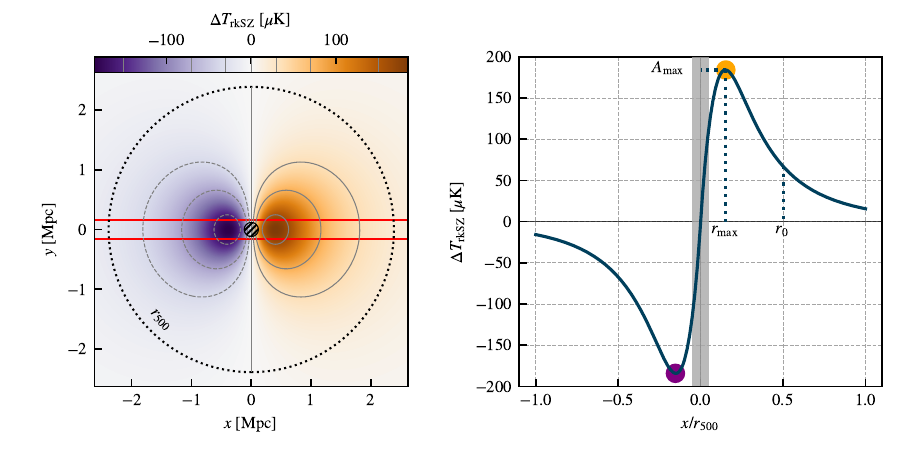}
    \caption{Model of the rkSZ signal map (left) and equatorial slice (right) based on the MACSIS 0 cluster at $z=0$. The resolution of the map is 20.5 kpc/pixel (256 pixels) and the horizontal red slice has a width of 657 kpc (32 pixels). In the left panel, we show line contours for $\Delta T_{\rm rkSZ}=0~{\rm \mu K},\pm 30 ~{\rm \mu K},\pm 70~{\rm \mu K},\pm 150~{\rm \mu K}$. We also indicate $r_{500}$ with a dotted circle. The inner hatched circle represents the $0.05\,r_{500}$ region excluded in the analysis. In the right panel, this region is represented as a vertical grey band between $\pm 0.05\,r_{500}$. The maximum (and minimum) rkSZ amplitude is $A_{\rm max}$, occurring at a radius $r_{\rm max}$. We also show the position of $r_0$ as a guideline, and we mark the extreme values in the profile with the same colours as in the map.}
    \label{fig:rksz_model}
\end{figure*}

The MACSIS hydrodynamic simulations \citep{macsis_barnes_2017} are a suite of 390 galaxy clusters simulated with dark matter, gas and sub-grid physics using the \textsc{Gadget-3} smoothed-particle hydrodynamics (SPH) code \citep[updated from \textsc{Gadget-2}, see][for details]{2005MNRAS.364.1105S}. These objects were initially selected from a (3.2 Gpc)$^3$ dark-matter-only parent volume based on their friends-of-friends (FoF) mass ($10^{15}<M_{\rm FoF}/{\rm M}_\odot<10^{16}$) at redshift $z=0$. The FoF groups within this mass range were then placed in logarithmic mass bins with a constant width of 0.2 dex. The two largest mass bins contained 7 and 83 halos, while 100 halos were selected from each of the three lowest mass bins\footnote{The mass bins with $\log_{10}(M_{\rm FoF}/{\rm M}_\odot)$ in the interval 15.0-15.2, 15.2-15.4 and 15.4-15.6 were further divided into 10 sub-bins each (constant logarithmic spacing of 0.02 dex), and 10 halos were selected from each sub-bin, amounting to 100 halos per bin. This method was performed to minimise bias selection towards low masses, caused by the steep slope of the halo mass function.}. The 390 halos were then re-simulated individually using the zoom-in technique \citep{1993ApJ...412..455K, 1997MNRAS.286..865T}, firstly in dark-matter-only mode and then with full physics. The MACSIS project used the same sub-grid model, particle-mass resolution and softening as in the BAHAMAS project \citep{2017MNRAS.465.2936M}: for the hydro-simulations, the dark matter particles had a mass of $6.49 \times 10^9$~M$_\odot$ and the gas particles had an initial mass of $1.18 \times 10^9$~M$_\odot$. \cite{macsis_barnes_2017} showed that the MACSIS (and the combined BAHAMAS+MACSIS) sample are in good agreement with the mass dependence of the observed hot gas fraction, the X-ray luminosity and the SZ Compton-$y$ parameter at $z=0$. To define the halos and substructures, we use the products of the SUBFIND code \citep{2001MNRAS.328..726S, 2009MNRAS.399..497D}. We also define galaxies as self-bound substructures with an associated stellar mass contained in a 70 kpc-radius spherical aperture above $10^{10}$ M$_\odot$, as in \cite{2019MNRAS.484.1526A}. The selection rules used in the definition of the MACSIS sample lead to an underrepresented low-mass halo population, biased towards low concentrations and high dark-matter spin parameter \citep{2017MNRAS.465.3361H}. Since our work focuses on the rotational dynamics of the cluster population, we discard MACSIS clusters below $M_{200} = 10^{14.5} h^{-1}$ M$_\odot$ at $z=0$ to mitigate this bias\footnote{$M_{200}$ is defined as the total mass within a spherical overdensity of radius $r_{200}$, centred in the gravitational potential minimum. $r_{200}$ is the radius at which the internal mean density exceeds the critical density by a factor of 200. Similarly, this approach is used to define $r_{500}$ and $M_{500}$.}. We further discard a small number (13) of clusters with abnormally low $f_{\rm gas}<0.05$ at $z=0$, likely caused by an AGN feedback event at high redshift. After the selection, the MACSIS sample is reduced from 390 to 377 clusters.

In Fig.~\ref{fig:cluster_display}, we show the most massive cluster in the sample, MACSIS 0 ($M_{500}=3.6\times10^{15}$ M$_\odot$), at $z=0$ with hot gas and stellar components. The map in the background encodes the gas density along the LoS and the star particles are shown as yellow markers in the foreground.
{In bottom-right corner of Fig.~\ref{fig:cluster_display}, we show} a map of the kSZ amplitude without the bulk motion. {This highlights the contribution from cluster rotation to the kSZ morphology, {which should become strongly visible when added to the approximately null tSZ signal at $\nu\simeq 217\,{\rm GHz}$} (CM02 and upper left corner of Fig.~\ref{fig:cluster_display}).} We note that for this particular projection, the bulk velocity of the cluster is perpendicular to the LoS, and only a smaller radial component boosts the negative part of the rkSZ dipole. When the bulk velocity is oriented radially, its contribution to the kSZ signal is larger, as in the visualisation of MACSIS 100 in Fig. \ref{fig:cluster_display_100}. Using this comparison, we stress that the kSZ amplitude strongly depends on the geometry of the observer, the cluster bulk velocity and bulk angular momentum.

\subsection{Modelling the rkSZ signal}
\label{sec:rksz-modelling}
{We will now describe the technique used to obtain this projected map and introduce the formalism that provides a analytic model for the rkSZ profiles.}
The kSZ Compton-$y$ parameter, given by the temperature difference $\Delta T_{\rm kSZ}$ over the CMB blackbody temperature, $T_{\rm CMB}$, is proportional to the mass-weighted velocity integrated along the LoS \citep{ksz_sunyaev_1980, mroczkowski2019_review}:
\begin{equation}
\label{eq:kSZ_definition}
     y_{\rm kSZ} \equiv \frac{\Delta T_{\rm kSZ}}{T_{\rm CMB}} = -\frac{\sigma_T}{c}\int_{\rm LoS} n_e \mathbf{v}\cdot {\rm d}\mathbf{l}.
\end{equation}
Here, $n_e$ is the free-electron number density and the product $\mathbf{v}\cdot d\mathbf{l}$ is the radial component of the velocity of the electron cloud element.
To construct a map of the projected Compton-$y$ parameter from SPH simulations, we discretise the integral above and sum the contributions of each particle $j$ to pixel $i$ as
\begin{equation}
    y_{{\rm kSZ,}i} = -\frac{\sigma_T}{c\, m_{\rm P}\, \mu_{\rm e}} \sum_{j} m_j v_{{\rm LoS,}j}\, W_{ij}(h_j),
\end{equation}
where $m_{\rm P}$ is the mass of the proton, $\mu_{\rm e}=1.14$ is the mean molecular weight per free electron, $v_{{\rm LoS,}j}$ is the velocity along the LoS, $h_j$ is the SPH smoothing length of the gas particle $j$ and $W_{ij}$ is the Wendland-C2 kernel \citep{wendland1995piecewise}, as implemented in \textsc{SWIFTsimIO} \citep{Borrow2020}\footnote{The calculation of the smoothed projection maps is performed using the \textsc{SWIFTsimIO} \textit{subsampled} backend, which guarantees converged results by evaluating each kernel 32 times or more. The overlaps between pixels are taken into account for every particle \citep[see][for further details]{Borrow2021}}.

We now derive the scaling relation of the kSZ amplitude with mass and redshift, predicted from self-similar cluster properties. Assuming fully ionised primordial gas, we obtain that $n_e$ is proportional to the critical density of the Universe, 
\begin{equation}
    \rho_{\rm crit}(z) = E^2(z)\, \frac{3 H_0^2}{8 \pi G},
\end{equation}
where $E^2(z)\equiv H^2(z) / H_0^2 =\Omega_{\rm m}(1+z)^3 + \Omega_\Lambda$, yielding $n_e \propto E(z)^2$. Similarly, assuming that the motion scales with the circular velocity of the cluster, we have 
{$|\mathbf{v}| \propto E(z)^{1/3}\, M_{500}^{1/3}$} 
and the electron column along the line-of-sight $\int |d\mathbf{l}| \propto r_{500} \propto  M_{500}^{1/3}\, E(z)^{-2/3}$. By combining these scaling relations in {Eq.~\eqref{eq:kSZ_definition}}, we obtain the predicted scaling for the kSZ amplitude:
\begin{equation}
    y_{\rm kSZ} \propto \Delta T_{\rm kSZ} \propto M_{500}^{2/3}\, E(z)^{5/3},
    \label{eq:kSZ-scaling}
\end{equation}
which is expected to moderately increase with cluster mass, and significantly increase with redshift. We note that this relation refers to the Compton-$y$ measured along the LoS, and not the Compton-$y$ integrated over the solid angle of the cluster. Compared to {the tSZ scaling} ($y_{\rm tSZ} \propto \int n_e T\, {\rm d}l \propto M_{500}\, E(z)^{2}$), the kSZ scaling has a weaker dependence on both mass and redshift.

{Assuming spherical symmetry and} following the formulation of \cite{2017MNRAS.465.2584B} for the rotating ICM, $\Delta T_{\rm kSZ}$ can be expressed in polar coordinates $(R, \phi)$ by the integral
\begin{equation}
\label{eq:kSZ-projection}
     y_{\rm rkSZ}(R,\phi) =-\frac{\sigma_T}{c}R\cos \phi \sin i \int_{R}^{r_{500}} n_e(r)~\omega(r)~\frac{2r~{\rm d}r}{\sqrt{r^2-R^2}},
\end{equation}
{with $R\in [0, r_{500}]$}. Here, the radial electron number density profile $n_e(r)$ and the angular velocity profile $\omega(r)$ are clearly separated in the integrand\footnote{The $\frac{2r}{\sqrt{r^2-R^2}}$ part of the integrand is a projection factor computed by applying a direct Abell transform. The integration limits also reflect this mapping.}. The inclination angle $i$ has the effect of reducing the overall amplitude of the rkSZ effect for $|\sin i|<1$.
This model shows that the rkSZ pattern is fully described by the choice of number density and angular velocity profiles. In our study, we model the rkSZ signal by assuming a \cite{2006ApJ...640..691V} $n_e(r)$ model {with the functional form}:
\begin{equation}
n_e= n_0\, \frac{\left(r / r_{\rm c}\right)^{-\alpha / 2}}{\left(1+r^{2} / r_{\rm c}^{2}\right)^{3 \beta / 2-\alpha / {4}}} \frac{1}{\Big(1+{[r/r_{\rm s}]^\gamma}\Big)^{\varepsilon / {2}\gamma}},
\label{eq:V06:emissionmeasure}
\end{equation}
where $\gamma = 3$ and $\varepsilon < 5$ are constrained and the other parameters are free to vary in the positive real interval.  {Following \cite{2017MNRAS.465.2584B} we found that the angular rotational velocity profile is well matched to the MACSIS data by}
\begin{equation}
\label{eq:omega-profile}
    \omega(r) = \frac{v_{\rm t0}}{r_0 \left[1 + (r/r_0)^\eta\right]},
\end{equation}
with the velocity scale radius $r_0$, the tangential velocity scale $v_{\rm t0}$, and the dimensionless slope parameter $\eta$ are to be determined by fitting the profile. Eq.~\eqref{eq:omega-profile} is a generalisation of the profile used to describe the rkSZ profiles of six relaxed MUSIC clusters \citep[fixing $\eta = 2$,][]{2017MNRAS.465.2584B}. Since most MACSIS clusters are dynamically unrelaxed and span over a wide range of halo masses, we allow $\eta$ to vary in the range $[1, 3]$ to match the slope of the decaying profile outside the peak radius. We illustrate the effect of changing the $r_0$, $v_{\rm t0}$ and $\eta$ parameters on the angular velocity and rkSZ profiles in Appendix~\ref{app:profile-fits}.

In Fig. \ref{fig:rksz_model}, we show an example an rkSZ map (top) modelled on the same cluster as above. For the density profile, we fit the $n_e(r)$ profile in Eq.~\eqref{eq:V06:emissionmeasure} with best-fit parameters
$$\left\{\frac{n_0}{\rm cm^{-3}}, \frac{r_{\rm c}}{r_{500}}, \frac{r_{\rm s}}{r_{500}}, \alpha,  \beta, \varepsilon\right\} = \{3.7 \times 10^{-3}, 0.17, 0.75, 1.5, 0.59, 2\}.$$
{We provide details on the profile fitting strategy in Section \ref{sec:results:fitting}.}

For the $\omega(r)$ profile, we assume a tangential velocity scale equal to the circular velocity of the cluster at $r_{500}$, given by $v_{\rm t0} \simeq v_{\rm circ} \simeq \sqrt{G M_{500}/r_{500}}$ and $r_0=r_{500} / 5$. We use the $n_e(r)$ fit from MACSIS~0 just to construct an example of rotation map and we do not fit for $v_{\rm t0}$ and $r_0$ at this stage. The top panel of Fig. \ref{fig:rksz_model} also shows the $r_{500}$ radius, and in solid grey the $0.05~r_{500}$ radius excluded from the \cite{2006ApJ...640..691V} fit. This threshold radius was determined by the particle-softening scale (4 $h^{-1}$ physical-kpc at $z < 3$) normalised to the value of $r_{500}$ for the smallest cluster in the MACSIS sample at $z=0$. To measure the amplitude of the rkSZ effect, we consider a horizontal slice through the centre of the halo and we average the pixel values in each column to give the rotation profile in the bottom panel of Fig. \ref{fig:rksz_model}. There, we also indicate the radius excluded from the density profile fit and the $r_{500}$ radius.

\section{Cluster properties and alignment between halo components}
\label{sec:cluster-properties}
Massive clusters form deep gravitational potential wells, causing the gas density to increase towards the centre. We therefore expect cluster atmospheres in large systems to produce more intense {SZ} signals. Moreover, MACSIS-like clusters acquire mass through a long history of mergers and accretion of matter from the surrounding filaments. For this reason, they present a complex dynamics even at $z=0$, making them ideal for future \textit{blind} rkSZ detection, {i.e. without prior knowledge of the cluster rotation axis}. In this paper, not only do we study the dependence of the rkSZ signal amplitude to the halo mass, $M_{500}$, but we also investigate its variations {with} other cluster properties, which we group in two categories: basic cluster properties and metrics for dynamical state. We will now investigate the MACSIS data set using these two classes of metrics and their correlations. 

\begin{figure*}
    \centering
	\includegraphics[width=2.05\columnwidth]{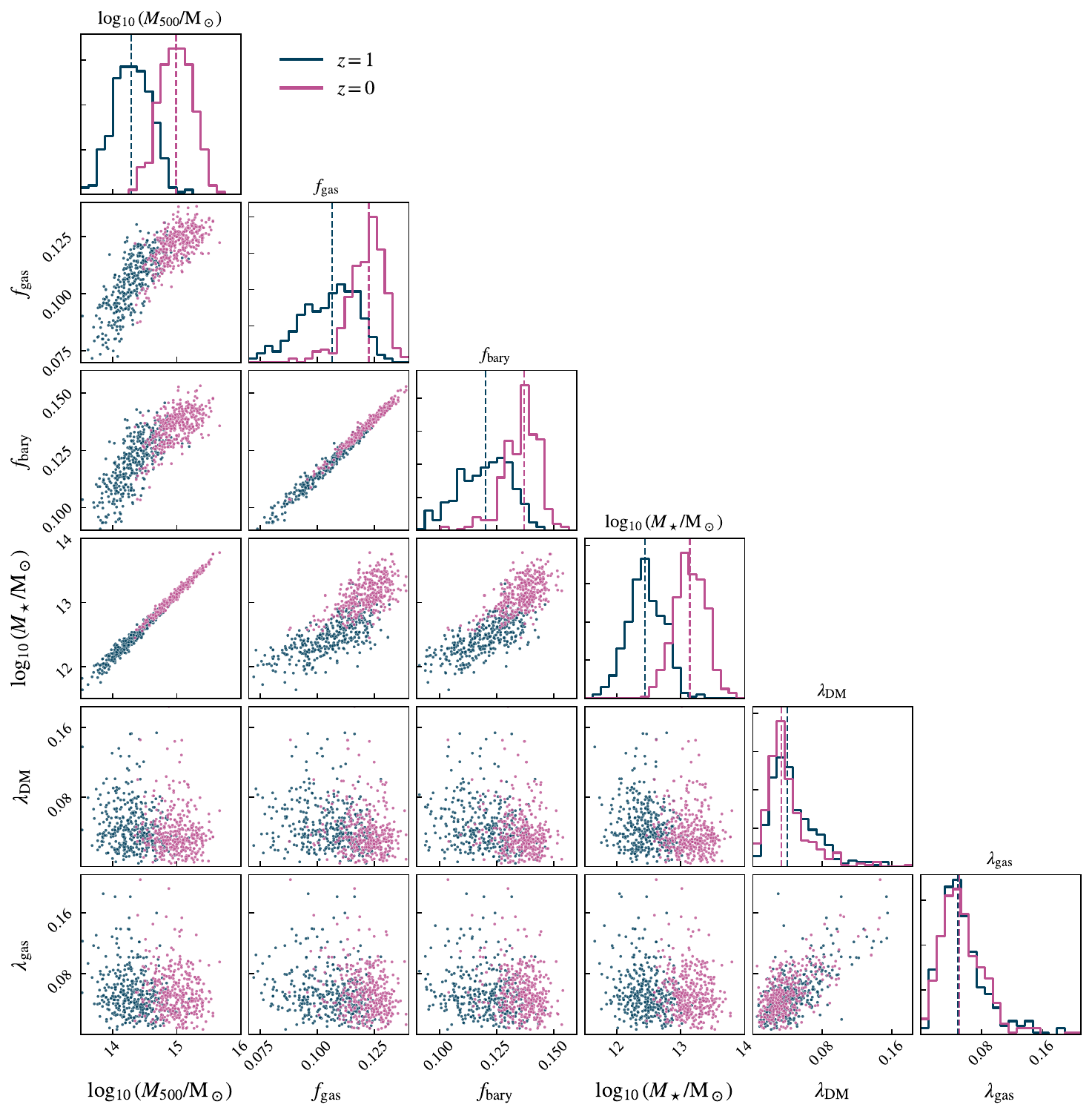}
    \caption{Corner plot with distributions of the basic cluster properties at $z=0$ (pink) and $z=1$ (blue). From left to right and top to bottom, we show the halo mass $M_{500}$, the hot gas fraction $f_{\rm gas}$, the baryon fraction $f_{\rm bary}$, the stellar mass $M_\star(r<r_{500})$, the dark matter spin parameter $\lambda_{\rm DM}$ and the hot gas spin parameter $\lambda_{\rm gas}$. Along the principal diagonal, we show the p.d.f. histograms of the cluster population, after removing the anomalous objects as specified in Section \ref{sec:datasets}. In each histogram, we indicate the sample median with vertical dashed lines. For clarity, each histogram in the diagonal plots is shown with a title indicating the quantity represented by the x-axis.}
    \label{fig:corner-plot-basic-properties}
\end{figure*}

\subsection{Basic cluster properties}
\label{sec:alignment:properties}
We begin by defining the global cluster properties used in the rest of this analysis. The hot gas fraction is $f_{\rm gas}\equiv M_{\rm gas}(r<r_{500})/M_{500}$, where $M_{\rm gas}(r<r_{500})$ is the total mass of the gas above the hydrogen ionisation temperature (we apply a temperature cut of $T\geq 10^5$ K) inside $r_{500}$. Similarly, we define the star fraction $f_\star \equiv M_\star(r<r_{500})/M_{500}$, where $M_\star(r<r_{500})$ is the stellar mass in $r_{500}$, and hence the baryon fraction $f_{\rm bary} = f_{\rm gas} + f_\star$.

The dark matter spin parameter, $\lambda_{\rm DM}$, measures the fraction of mechanical energy of the clusters due to rotation and is estimated using the relation by \cite{2001ApJ...555..240B}:
\begin{equation}
    \label{eq:spin-parameter}
    \lambda_{\rm DM} = \frac{j_{\rm DM}}{\sqrt{2} v_{\rm circ} r_{500}},
\end{equation}
where $j_{\rm DM}$ is the specific angular momentum (about the centre of potential) of the dark matter in $r_{500}$ and $v_{\rm circ}=\sqrt{GM_{500}/r_{500}}$ the circular velocity at $r_{500}$. We use $\lambda_{\rm DM}$ to classify slow and fast dark-matter rotators in the MACSIS sample. {In analogy to $\lambda_{\rm DM}$,} $\lambda_{\rm gas}$ instead uses the specific angular momentum of the hot gas $j_{\rm gas}$ to quantify the fraction of kinetic energy of the ICM associated with the bulk rotation. 

In Fig. \ref{fig:corner-plot-basic-properties}, we show these quantities for the MACSIS clusters at $z=0$ (pink) and $z=1$ (blue) in a corner plot, with the probability density functions (p.d.f.s) along the principal diagonal. For each p.d.f., a vertical dashed line indicates the median value, which we will use to split the $z=0$ sample in Section \ref{sec:results}. The $M_{500}$ p.d.f. on the top-left corner is the normalised halo mass function (HMF), whose shape is determined by the sample selection method of \cite{macsis_barnes_2017}. We will discuss the MACSIS HMF further in Section \ref{sec:results:z-dependence}, when selecting clusters to compare the $z=0$ and $z=1$ populations. Below the $M_{500}$ p.d.f., the $f_{\rm gas}$-$M_{500}$ and $f_{\rm bary}$-$M_{500}$ panels show the hot gas and baryon mass-scaling relations. These reproduce the results shown by \cite{macsis_barnes_2017}. The p.d.f. for $\lambda_{\rm DM}$ and $\lambda_{\rm gas}$ show a population distribution compatible with early results by \cite{2001ApJ...555..240B} and, in addition, suggest that $\lambda_{\rm gas}$ is correlated with $\lambda_{\rm DM}$, but both are only weakly correlated with the other basic cluster properties. Neither the $\lambda_{\rm gas}$ or $\lambda_{\rm DM}$ distributions change significantly with redshift.

\begin{figure*}
    \centering
	\includegraphics[width=2.05\columnwidth]{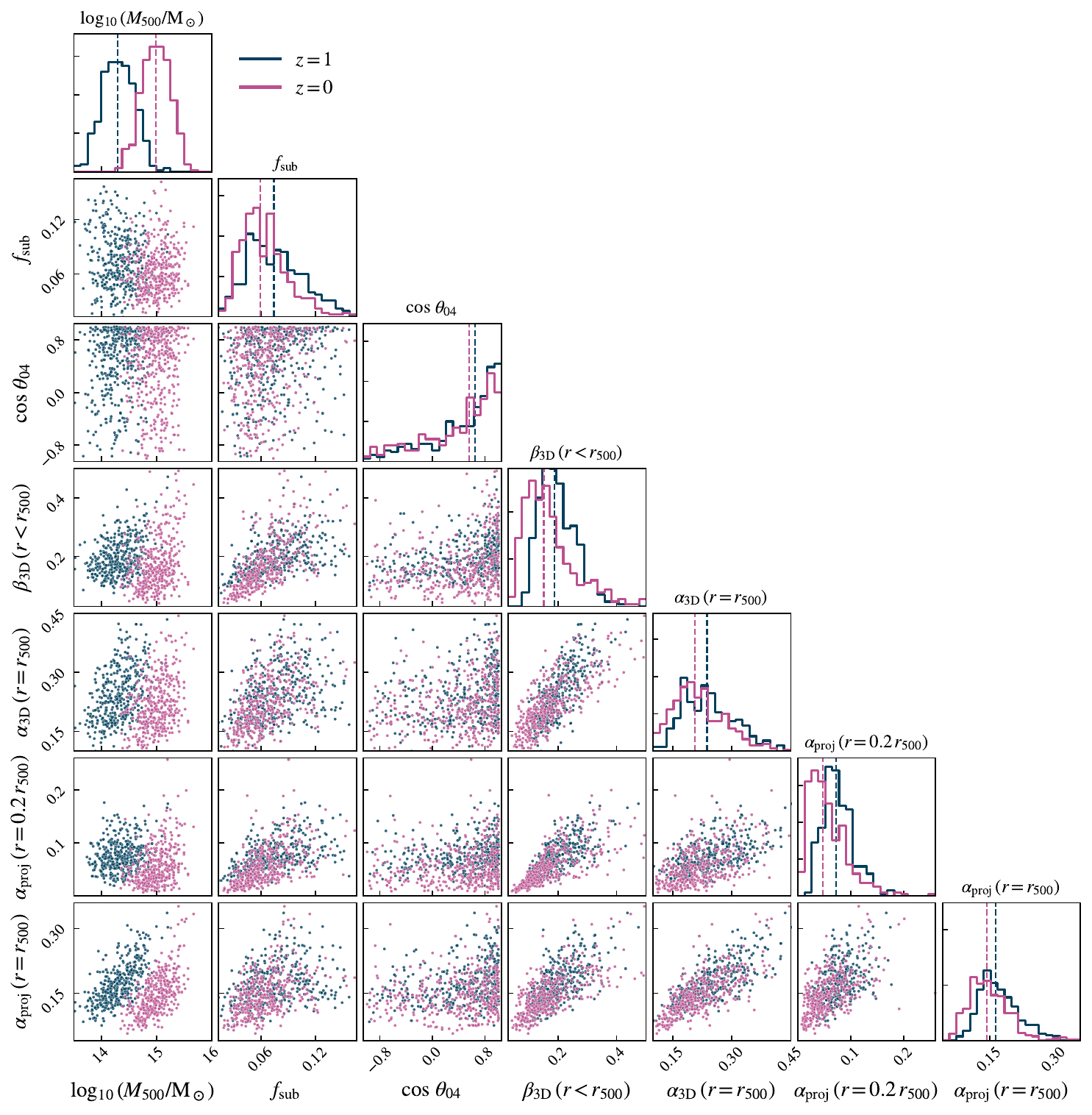}
    \caption{As in Fig. \ref{fig:corner-plot-basic-properties}, but focusing on the dynamical state metrics: the substructure mass fraction $f_{\rm sub}$, the angle between the angular momenta of the hot gas and the galaxies $\cos\,\theta_{04}$, the overall kinetic-to-thermal ratio $\beta_{\rm 3D}$ inside $r_{500}$, the 3D non-thermal pressure fraction at $\alpha_{\rm 3D}$ evaluated at $r_{500}$ and, finally, the projected non-thermal pressure fraction $\alpha_{\rm proj}$ evaluated at in the core ($0.2\,r_{500}$) and at at $r_{500}$.}
    \label{fig:corner-plot-dynamical-state}
\end{figure*}

\subsection{Metrics for dynamical state and spin geometry}
\label{sec:alignment:spin}
To quantify how thermodynamically perturbed the ICM is, we use the kinetic-to-thermal (energy) ratio $\beta \equiv E_{\rm kin}/E_{\rm th}$ as a proxy. Here, $E_{\rm kin}=1/2 \sum m_i ({\bf v}_i - {\bf v_{\rm bulk}})^2$ is the total kinetic energy of the hot gas within $r_{500}$, computed from the mass $m_i$ and the velocity ${\bf v}_i$ of the particles after subtracting the bulk velocity ${\bf v_{\rm bulk}}$; assuming all gas is mono-atomic, the thermal energy $E_{\rm th}=3/2\, {\rm k_B}\sum T_i N_{{\rm H},i}$, where $T_i$ is the temperature of the gas particle, the number of hydrogen atoms is $N_{{\rm H},i}=m_i/(\mu\,m_{\rm P})$, $\mu\simeq 1.16$ is the mean molecular weight and $m_{\rm P}$ is the proton mass. Large values of the kinetic-to-thermal ratio ($\gtrsim 0.1$), typically measured in high-mass systems, are associated with a thermodynamically perturbed cluster atmosphere \citep[see e.g.][]{macsis_barnes_2017}. Throughout this work, $\beta_{\rm 3D}$ indicates the kinetic-to-thermal ratio computed for all hot gas particles in a 3D $r_{500}$ aperture.

From Fig. \ref{fig:corner-plot-dynamical-state}, we learn that 50\% of the MACSIS sample has a kinetic-to-thermal ratio lower than 0.15. However, we identify a handful of objects with $\beta_{\rm 3D}\approx0.5$. The hot gas in these systems have a particularly high kinetic energy (in the centre of potential, CoP, rest frame), which must be in the form of bulk rotation, internal bulk motions, and turbulence. The kinetic-to-thermal implicitly combines these two contributions. However, we can directly probe the kinetic energy in the rotational mode using $\lambda_{\rm gas}$, as defined above, and we use the non-thermal pressure, $P_{\mathrm{nth}}$, to quantify the kinetic energy in the form of turbulence and small-scale bulk motion \citep[e.g.][]{2018MNRAS.481L.120V}. After introducing the total (thermal plus non-thermal) pressure as $P_{\mathrm{tot}}$, we write the non-thermal support fraction as
\begin{equation}
    \alpha = \frac{P_{\mathrm{nth}}}{P_{\mathrm{tot}}} = \left[ 1 + \frac{3 k_{\rm{B}} T}{\mu m_{\rm{H}} \sigma_{\rm 3D}^2} \right]^{-1} = \frac{\beta_{\rm 3D}}{1 + \beta_{\rm 3D}}.
    \label{eq:alpha}
\end{equation}
In this expression, we use the 3D velocity dispersion, $\sigma_{\rm 3D}^2$, following \cite{2022arXiv221101239T}. In analogy to their approach, we adopt two methods, which we summarise below.
\begin{enumerate}
    \item {\bf Radial $\alpha_{\rm 3D}$ profiles}. We first compute the $j \in \{x, y, z\}$ components of the velocity dispersion along the three axes of the simulation box: 
    \begin{equation}
        \sigma_{j}^{2} = \frac{\sum_{i} W_{i} \left( v_{i,j} - \bar{v}_{j} \right)^{2}}{\sum_{i} W_{i}}.
        \label{eq:velocity-dispersion}
    \end{equation}
    Here, $\bar{v}$ is the bulk velocity of ensemble of particles $\{i\}$ within a spherical shell at radius $r$, and $W$ is a weighting function which in this case is set to the particle mass, $W_i = m_i$. Then, we find the total velocity dispersion by adding the components in quadrature as $\sigma_{\rm 3D}^2 = \sum_j \sigma_{j}^{2} \equiv \sigma_{x}^{2} + \sigma_{y}^{2} + \sigma_{z}^{2}$. This value is substituted in Eq.~\eqref{eq:alpha} to give the $\alpha_{\rm 3D}(r)$ profile. In this work, we use the value of $\alpha_{\rm 3D}(r = r_{500})$ as a metric for the turbulence derived from 3D profiles. We stress that $\alpha_{\rm 3D}$ computed in this manner is independent of the choice of LoS. Moreover, $\alpha_{\rm 3D}$ measures the \textit{local} value at $r_{500}$, unlike $\beta_{\rm 3D}$, which is an \textit{integrated} quantity.
    
    \item {\bf Projected $\alpha_{\rm 2D}$ profiles}. In our study, we employ an additional metric to estimate the non-thermal support fraction following the prescription of \cite{Roncarelli2018MeasuringObservations}. In Eq.~\eqref{eq:velocity-dispersion}, we use the X-ray emission measure as a weighing function, $W_{i}=\rho_i m_i$, and we construct one estimate of the total velocity dispersion for each axis of the simulation volume: $\sigma_{{\rm 3D}, j}^2 = 3\, \sigma_{j}^2$. These results lead to three projected profiles, $\alpha_{{\rm proj}, j}(r)$, estimated along perpendicular LoS's. As in the previous method, we aim to construct an direction-agnostic formulation of $\alpha$. We achieve this by averaging the projected $\alpha$ profiles along the three axes as $\alpha_{\rm proj}(r) = 1/3 \, \sum_j \alpha_{{\rm proj}, j}(r)$. Finally, we evaluate the averaged projected profile at $0.2\, r_{500}$ (in the core) and $r_{500}$, to obtain $\alpha_{\rm proj}(r = 0.2\, r_{500})$ and $\alpha_{\rm proj}(r = r_{500})$ respectively. The $\alpha_{\rm 2D}$ profiles are computed as the median of 12 azimuthal bins. This method was shown to be more effective in reducing local fluctuations then the spherical average method \citep[see e.g. Section 3.2 of][]{2022arXiv221101239T} used in the derivation of the $\alpha_{\rm 3D}$ profiles.
\end{enumerate} 
In Eq. \eqref{eq:alpha}, we emphasize that $\alpha$ can be written as a function of $\beta$. This connection exists because: (i) $({\bf v}_i - {\bf v_{\rm bulk}})^2$ in $E_{\rm kin}$ can be expressed in terms of the velocity dispersion $\sigma^2$ in $P_{\rm nth}$ and (ii) the (mass-weighted) temperature in $E_{\rm th}$ is the same used in the \textit{thermal} pressure fraction, which gives $P_{\rm tot}$. A detailed derivation is given in Appendix \ref{app:correlation:alpha-beta}.
To assess the dynamical state of a cluster, we also use the substructure fraction $f_{\rm sub}$, defined as the fraction of the mass in a FoF group bound to substructures inside $r_{500}$. For instance, \cite{2017MNRAS.465.3361H} use $f_{\rm sub}$ to classify relaxed clusters if $f_{\rm sub} < 0.1$. Crucially, substructures falling into the cluster's potential well are known to transfer angular momentum into the system and perturb the ICM, potentially enhancing the kSZ amplitude due to cluster rotation.

We show the {distribution} of these properties for the MACSIS sample in Fig. \ref{fig:corner-plot-dynamical-state}. We recover positive correlation between $\beta_{\rm 3D}$ and $f_{\rm sub}$ as expected and we also show a similar correlation between the three estimates for $\alpha$ at $z=0$ and $z=1$ with $M_{500}$, $\beta_{\rm 3D}$ and $f_{\rm sub}$. {None of the dynamical state metrics show very strong dependence on redshift.}

Next, we illustrate the framework for computing $\cos\theta_{04}$, {a metric for the alignment of the gas and galaxies spins,} and discuss its value in the context of the MACSIS clusters. The orientation of the angular momentum of the cluster components (gas, dark matter and stars) relative to each other is important when scaling, {reorienting} and stacking kSZ maps, for instance in the study by \cite{2019JCAP...06..001B}. In this work, we mimic their analysis method by computing the specific angular momentum (or simply \textit{spin}) of each cluster component as ${\bf j}_k = {\bf J}_k / M_k$ with $k\in\{{\rm gas,~DM,~stars}\}$, $M_k$ the component mass in $r_{500}$ and the angular momentum
\begin{equation}
    {\bf J}_k = \sum_{i:r<r_{500}} m_i ~ \left({\bf v_i} - {\bf v_{\rm bulk}}\right) \times \left({\bf r_i} - {\bf r_{\rm CoP}}\right) \Bigg|_{k\in\{{\rm gas,~DM,~stars}\}}.
\end{equation}
Note that ${\bf J}_k$ is computed about the centre of potential at position ${\bf r_{\rm CoP}}$ and in the cluster's rest frame, obtained by subtracting the bulk velocity ${\bf v_{\rm bulk}}$. Hereafter, we re-label the cluster components with the \texttt{ParticleType} notation used in \textsc{Gadget-3} as follows $k\in\{{\rm gas,~DM,~stars}\}\longrightarrow \{0, 1, 4\}$. We then compute the angle $\theta$ between the spin vectors of the components $(m,n) = (0,1), (0,4), (1,4)$ as
\begin{equation}
    \cos \theta_{mn} = \frac{{\bf j}_m \cdot {\bf j}_n}{|\, {\bf j}_m\, |\, |\, {\bf j}_n\, |},
    \label{eq:dotproduct}
\end{equation}
as also done by, e.g., \cite{2002ApJ...576...21V, 2010MNRAS.404.1137B} and \cite{2017MNRAS.466.1625Z}. From Eq.~\eqref{eq:dotproduct}, a value of $\cos \theta_{mn}\approx 1$ indicates that the spins are aligned, while anti-aligned spins return $\cos \theta_{mn}\approx -1$. 
For the calculation of the stellar spin, we only select the star particles in galaxies, following the definition in Section \ref{sec:datasets}.
While we only show data for $\cos \theta_{04}$ in Fig. \ref{fig:corner-plot-dynamical-state}, we report the other combinations $(m, n)$ combinations in Fig.~\ref{fig:corner-plot-all-properties}. 

We {find that, in all $\cos \theta_{mn}$ histograms,} most of the clusters have components with well-aligned spins, as expected for structures forming in a $\Lambda$CDM universe. We obtained sharply peaked distributions for $\cos \theta_{(01), (14)}\approx 1$, suggesting that the galaxies and the hot gas co-rotate with the DM halo, which accounts for the largest mass content (and angular momentum) in a cluster. The spins of the gas and the galaxies, however, are more poorly aligned with each other when compared to the (01) or the (14) pairs, shown by a broader tail in the distribution. The de-rotation method in \cite{2019JCAP...06..001B} uses the spin from the galaxies as proxy for the gas spin, however, our results prove that a MACSIS-like cluster population may have a large fraction of objects where this assumption is not valid and may affect the rkSZ amplitude in the stacked maps. We will corroborate this claim in Section \ref{sec:image-processing:stacking}. 

{We find no correlation between the alignment angles,
$\{ \cos \theta_{01}$, $\cos \theta_{04}$, $\cos \theta_{14} \}$, and the 
dynamical state indicators, 
$\{ \beta_{\rm 3D}$, $f_{\rm sub}$, $\alpha_{\rm 3D}$, $\alpha_{\rm proj}\}$, 
as illustrated by the $\cos \theta_{04}$ data in  Fig. \ref{fig:corner-plot-dynamical-state}}. The $\cos \theta_{mn}$ quantities are 
{also} not found to be correlated with the halo mass or the hot gas fraction, while they are positively correlated with each other. These results suggest that the tail of the $\cos \theta_{04}$ distribution is equally represented at all halo masses in the MACSIS sample. The effects of the misalignment of the galaxies and the hot gas spins are therefore expected to statistically affect the stacking of the rkSZ maps equally across the MACSIS mass range if the orientation of ${\bf j}_4$ is used as a proxy for that of ${\bf j}_0$. From Fig. \ref{fig:corner-plot-dynamical-state}, we find $\cos \theta_{04} < 0$ in about 20\% of the cluster sample. For these objects, the angle between the angular momenta of gas and galaxies is very large $>90^\circ$, with a few cases reaching $\approx 180^\circ$ (i.e. ${\bf j}_0$ and ${\bf j}_4$ are anti-aligned). Coherent rotation, expressed by small $\theta_{04}$ is a well-established result of tidal-torque theory and finding indications of counter-rotating gas and galaxy components may seem puzzling. We find that MACSIS clusters with $\cos \theta_{04} < 0$ also have low values of $\lambda_{\rm DM}$ and $\lambda_{\rm gas}$ (see Fig.~\ref{fig:corner-plot-all-properties} in Appendix \ref{app:correlation-coefficients}). Since $\lambda \propto |{\bf j}|$, as shown in Eq.~\eqref{eq:spin-parameter}, cluster with incoherent rotation tend to have a low angular momentum in $r_{500}$. This scenario can occur if the angular momenta of the individual gas particles (and galaxies) are not well-aligned aligned, leading to a small vector sum. We tested this hypothesis on 5 clusters with low $\cos \theta_{04}$ values at $z=0$ and verified that the distribution of the individual angular momenta about the CoP was overall isotropic for the gas, and even more so for the galaxies. Bulk rotation clearly has a small impact on the overall dynamics in clusters with this characteristic, and the orientation of ${\bf j}_0$ and ${\bf j}_4$ may simply be dictated by small excess contributions of, e.g., a substructure entering $r_{500}$ or exerting a gravitational tidal torque from nearby.

In Appendix \ref{app:correlation-coefficients}, we include the complete corner plot with both basic properties and dynamical state indicators; we also compute the Spearman correlation coefficients as a quantitative measure of the correlation between quantities.

\section{Image processing}
\label{sec:image-processing}

\subsection{De-rotation}
\label{sec:image-processing:derotation}
To recover the elusive rotational kSZ signal from noisy observations, or from substructure-rich clusters with complex dynamics, we generate and stack projected maps of the rkSZ signal. Following the stacking method in \cite{2019JCAP...06..001B}, we first rotate the particle (or galaxies) positions and the velocity vectors such that ${\bf J}_{\rm k}$ aligns with the $z$-axis of the parent box, identified with the unit vector $\mathbf{\hat{z}}=(0, 0, 1)$. The rotation transformation is implemented using Rodrigues' rotation formula \citep[see derivations for an $SO(3)$ rotation group in e.g.][]{bauchau2003vectorial, dai2015euler}. We compute the rotation axis vector $\mathbf{q}$ by taking the outer product of the two vectors ${\bf J_{\rm k}}$ and $\mathbf{\hat{z}}$:
\begin{equation}
   \mathbf{q} \equiv \begin{pmatrix} q_x\\q_y\\q_z \end{pmatrix}=\frac{{\bf J_{\rm k}} \times \mathbf{\hat{z}}}{|{\bf J_{\rm k}} \times \mathbf{\hat{z}}|},
\end{equation}
and the angle between ${\bf J_{\rm k}}$ and $\mathbf{\hat{z}}$ given by the inner product $\cos \theta = {\bf J_{\rm k}} \cdot \mathbf{\hat{z}} / |{\bf J_{\rm k}}| = J_{{\rm k},z}/J_{\rm k}$.The skew-symmetric matrix is then defined as
\begin{equation}
\mathbf{Q}=
     \begin{bmatrix}
    0 & -q_z & q_y\\ q_z & 0 & -q_x\\ -q_y & q_x & 0,
    \end{bmatrix}
\end{equation}
and is used to compute the rotation matrix
\begin{equation}
\mathbf{R}=\mathbf{I}_3 + (\sin \theta) ~\mathbf{Q} + (1 + \cos \theta) ~\mathbf{Q}^2, 
\end{equation}
where $\mathbf{I}_3$ is the $3\times3$ identity matrix. We de-rotate the positions $\mathbf{r_i}$ and velocities $\mathbf{v_i}$ of the selected gas particles as follows
\begin{equation}
    {\bf r'}_i = \mathbf{R}^{-1}\, \left({\bf r}_i - {\bf r}_{\rm CoP}\right);\\
    {\bf v'}_i = \mathbf{R}^{-1}\, \left({\bf v}_i - {\bf v}_{\rm bulk}\right).
\end{equation}
By using the angular momentum of the gas, DM and galaxies, we construct three matrices $\mathbf{R}_k$ which align $\mathbf{J}_k$ with the $z$-axis of the box and produce different projections.

\begin{figure*}
	\includegraphics[width=2\columnwidth,trim=0.5cm 0.5cm 0cm 0cm,clip]{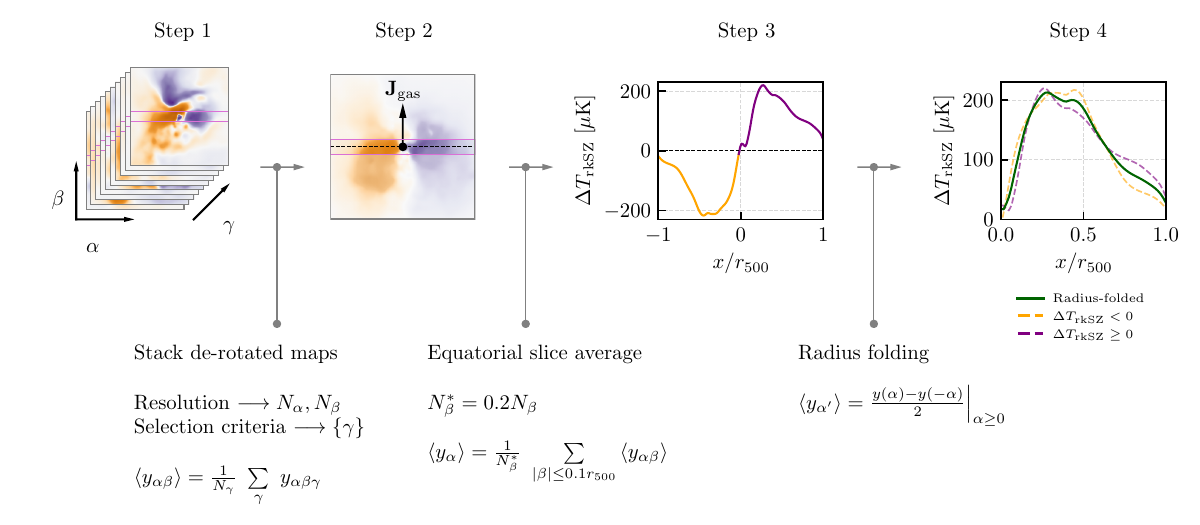}
    \caption{Workflow diagram for stacking rkSZ de-rotated and projected maps of MACSIS clusters, averaging the equatorial slice and folding the rotation profile to obtain the radial rkSZ profile. We define four steps and  The illustrations at the top show an example of this procedure by considering the rkSZ maps from the 10 most massive MACSIS clusters (step 1). In each map, the pixel coordinates are defined by the indices $(\alpha, \beta)$ and the index $\gamma$ runs over different maps. The 10 maps are averaged over $\gamma$ to produce the stacked map (step 2). Since these maps are constructed from an edge-on configuration, ${\bf J_{\rm gas}}$ points upwards in the plane of the map, as shown by the black arrow. From the stacked map, we select an equatorial slice (in red) of thickness $0.1\, r_{500}$, used to obtain the profile in step 3. Finally, this result is folded about the origin and averaged to produce the radial rkSZ profile in step 4. Throughout the diagram, we indicate positive Doppler boost in purple and negative in orange. In the right plot (step 4), the dashed lines are the positive and negative folded sections of the profile in step 3, and the solid line is their average.}
    \label{fig:stacking_workflow}
\end{figure*}

\subsection{Projections}
\label{sec:image-processing:projections}
We then compute the $y_{\rm rkSZ}$ contribution for each particle using the integrand in Eq.~\eqref{eq:kSZ_definition} and sum the contributions along the LoS to produce projected maps. In this work, we consider seven projections: 
\begin{enumerate}
    \item In the first three projections, $\{x,y,z\}$, the clusters are not de-rotated and the LoS aligned to each of the axes of the parent box. In $\Lambda$CDM cosmology, the orientation of cluster spins follows a random distribution for a large sample of objects; stacking maps without de-rotating is expected to smooth out the coherent dipole signature from cluster rotation, producing a noisy signal, {consistent with no rotation}.
    \item \textit{gas-edge-on} projection, where ${\bf J}_{\rm gas}$ is aligned and the LoS is perpendicular to the angular momenta via the $\mathbf{R}_{\rm gas}$ rotation matrix. This configuration is expected to produce the largest possible rotational signal.
    \item \textit{gas-face-on} uses the same alignment as above, but with the LoS perpendicular to the plane of rotation. This result is obtained by projecting the particle data \textit{along} $\mathbf{J}_z$, rather than perpendicularly to it. In this case, the stacked maps are not expected to show any rkSZ temperature signal.\footnote{At second order in the rotational speed an $E$-mode polarization pattern appears in this case \citepalias{CM02}, however, we neglect this contribution.}
    \item In the \textit{galaxies-edge-on} projection, the clusters are de-rotated to align the galaxies angular momenta and then viewed edge-on via the $\mathbf{R_{gal}}$ rotation matrix. This projection is designed to yield the largest signal if the galaxies are used as proxy for the gas angular momenta see \citep[see][]{2019JCAP...06..001B}.
    \item In the \textit{dark-matter-edge-on} projection uses the same prescription as \textit{galaxies-edge-on}, but the alignment is based on ${\bf J_{\rm DM}}$ and the $\mathbf{R_{DM}}$ rotation matrix.
\end{enumerate}
After obtaining ${\bf r'_i}$ and ${\bf v'_i}$ from the rotation matrices, we compute the rkSZ projection maps using the discretised form of Eq.~\eqref{eq:kSZ_definition}.

\subsection{Stacking and averaging}
\label{sec:image-processing:stacking}

After {orienting} the cluster datasets, we combine the rkSZ maps to produce the radial rotation profiles in four steps, summarised in Fig. \ref{fig:stacking_workflow}. In this example, we selected gas-aligned edge-on rkSZ maps from the 10 most massive MACSIS clusters. This sample defines the set $\{\gamma \}$, where $\gamma$ is an index running over the selected maps. The number of maps in the selection is $N_\gamma = 10$ in this demonstration. For each map, we specify the Cartesian position of pixels by an index pair $(\alpha, \beta)$, where $\alpha \in [0, N_\alpha]$,  $\beta \in [0, N_\beta]$ and $N_\alpha \times N_\beta$ is the resolution of the map. We stress that, in this section, $\alpha$ and $\beta$ are indices running over the pixels of the maps, and do not refer to the cluster properties. Using this set-up, we stack the maps by averaging over $\gamma$, to obtain the average map $\langle y_{\alpha \beta}\rangle$, as shown in Fig.~\ref{fig:stacking_workflow}. We consider an equatorial slice of thickness $0.2 \times r_{500}$ with $N^*_\beta = 0.2 \times N_\beta$ pixels, spanning the horizontal axis of the map, then we compute the  mean of the pixels in each pixel column. This operation leads to step 3 and produces the \textit{equatorial} rotation profile $\langle y_{\alpha}\rangle$. To facilitate the model fitting strategy, we further fold the negative section of the rotation profile onto the positive section via a reflection about the origin, $\langle y(\alpha<0) \rangle \longrightarrow -\langle y(-\alpha) \rangle$, and average the two sections to produce the radial rkSZ profile $\langle y_{\alpha '} \rangle \equiv y_{\rm rkSZ}(R)$.

As a final step in our analysis pipeline, we estimate the statistical scatter in the radial rkSZ profiles using the bootstrapping technique \citep{efron1979bootstrap, efron1987better}. We randomly sample the maps in the simulation catalogue with replacement and then we stack these to produce a mean profile (as in Fig. \ref{fig:stacking_workflow}). This operation is repeated $10^4$ times, producing as many profiles for each sample selection. From these $10^4$ realisations, we then compute the median profile, and define the uncertainties the difference between the first and third quartiles. We use this method for presenting the results from the MACSIS sample in Section \ref{sec:results}.

\section{Results}
\label{sec:results}

\begin{table*}
    \setlength{\tabcolsep}{5pt}
    \centering
    \caption{Summary of the selection criteria (column 1), the reference to the panel showing the radial rkSZ profile (column 2), the maximum measured rkSZ amplitude $A_{\rm max}$ and the peak-radius $r_{\rm max}$ for the gas-aligned profiles (columns 3 and 4), the galaxies-edge-on projection (columns 5 - 6) and the DM-edge-on projection (columns 8 and 9). In columns 7 and 10, we show the ratio between the maximum amplitude for the galaxies- and DM-aligned profiles with that of the gas-aligned profiles in the same sample of objects. These results are obtained at $z=0$ unless stated otherwise. To facilitate the visualisation of this table, we have highlighted the \textit{All clusters} row in grey; for the remaining rows, we alternate white and orange backgrounds to indicate the low- and high-quantity samples respectively. We will use the same colour coding in other tables throughout the paper.}
    \label{tab:max_measured}
    \rowcolors{9}{orange!20}{white}
    \begin{tabular}{lccccccccc}
    \toprule
     &
     \multicolumn{1}{c}{}    & 
     \multicolumn{2}{c}{\bf Gas-aligned} & 
     \multicolumn{3}{c}{\bf Galaxies-aligned} & 
     \multicolumn{3}{c}{\bf DM-aligned}\\ 
     
     \cmidrule(rl){3-4} \cmidrule(rl){5-7} \cmidrule(rl){8-10} \rule{0pt}{1ex}
     
      Selection criterion & Fig.  &  $A_{\rm max}$ & $r_{\rm max}$ &  $A_{\rm max}$ & $r_{\rm max}$ & Amplitude fraction & $A_{\rm max}$ & $r_{\rm max}$ & Amplitude fraction\\
                 & & [$\mu$K] & [$r_{500}$] & [$\mu$K] & [$r_{500}$] & $A_{\rm max}^{\rm (galaxies)} / A_{\rm max}^{\rm (gas)}$ &  [$\mu$K] & [$r_{500}$] & $A_{\rm max}^{\rm (DM)} / A_{\rm max}^{\rm (gas)}$ \\
    \midrule
    \rowcolor{gray!25}
    All clusters (377)                        &  \ref{fig:slices_projections}                                       &    80.2$\,\pm\,5.2$ & 0.20 & 32.1$\,\pm\,5.3$ & 0.20 & 0.40$\,\pm\,0.07$ & 50.1$\,\pm\,5.1$ & 0.20 & 0.63$\,\pm\,0.08$ \\
    $M_{500} < 9.7\times 10^{14}$ M$_\odot$   &  \cellcolor{white}                                                  &    32.5$\,\pm\,1.7$ & 0.23 & 12.0$\,\pm\,1.6$ & 0.28 & 0.37$\,\pm\,0.05$ & 20.7$\,\pm\,1.8$ & 0.24 & 0.64$\,\pm\,0.06$ \\
    $M_{500} > 9.7\times 10^{14}$ M$_\odot$   &  \cellcolor{white}\multirow{-2}{*}{\ref{fig:slices:properties}.A}   &    128.4$\,\pm\,10.8$ & 0.19 & 54.2$\,\pm\,10.7$ & 0.20 & 0.42$\,\pm\,0.09$ & 79.0$\,\pm\,9.8$ & 0.20 & 0.62$\,\pm\,0.09$ \\
    $f_{\rm gas}$ < 0.12                      &  \cellcolor{white}                                                  &    47.9$\,\pm\,3.4$ & 0.24 & 17.3$\,\pm\,3.2$ & 0.33 & 0.36$\,\pm\,0.07$ & 25.3$\,\pm\,2.5$ & 0.31 & 0.53$\,\pm\,0.06$ \\
    $f_{\rm gas}$ > 0.12                      &  \cellcolor{white}\multirow{-2}{*}{\ref{fig:slices:properties}.B}   &    108.7$\,\pm\,10.8$ & 0.18 & 49.2$\,\pm\,10.2$ & 0.18 & 0.45$\,\pm\,0.10$ & 78.5$\,\pm\,11.1$ & 0.13 & 0.72$\,\pm\,0.12$ \\
    $f_{\rm bary}$ < 0.12                      &  \cellcolor{white}                                                 &    55.1$\,\pm\,5.2$ & 0.24 & 22.5$\,\pm\,4.5$ & 0.27 & 0.41$\,\pm\,0.09$ & 29.9$\,\pm\,3.1$ & 0.28 & 0.54$\,\pm\,0.08$ \\
    $f_{\rm bary}$ > 0.12                      &  \cellcolor{white}\multirow{-2}{*}{\ref{fig:slices:properties}.C}  &    110.0$\,\pm\,11.1$ & 0.17 & 47.3$\,\pm\,9.9$ & 0.19 & 0.43$\,\pm\,0.10$ & 76.5$\,\pm\,10.7$ & 0.13 & 0.70$\,\pm\,0.12$ \\
    
    $M_{\star} < 9.7\times 10^{14}$ M$_\odot$   &  \cellcolor{white}                                                   &    31.7$\,\pm\,1.7$ & 0.22 & 11.7$\,\pm\,1.7$ & 0.26 & 0.37$\,\pm\,0.06$ & 21.8$\,\pm\,1.8$ & 0.23 & 0.69$\,\pm\,0.07$ \\
    $M_{\star} > 9.7\times 10^{14}$ M$_\odot$   &  \cellcolor{white}\multirow{-2}{*}{\ref{fig:slices:properties}.D}    &    128.8$\,\pm\,10.3$ & 0.20 & 53.7$\,\pm\,10.3$ & 0.20 & 0.42$\,\pm\,0.09$ & 79.1$\,\pm\,9.7$ & 0.20 & 0.61$\,\pm\,0.09$ \\
    
    $\lambda_{\rm DM}$ < 0.03                 &  \cellcolor{white}                                                     &    60.4$\,\pm\,3.9$ & 0.24 & 11.2$\,\pm\,3.2$ & 0.31 & 0.18$\,\pm\,0.05$ & 25.7$\,\pm\,5.4$ & 0.20 & 0.42$\,\pm\,0.09$ \\
    $\lambda_{\rm DM}$ > 0.03                 &  \cellcolor{white}\multirow{-2}{*}{\ref{fig:slices:properties}.E}      &    102.1$\,\pm\,12.4$ & 0.17 & 59.5$\,\pm\,8.9$ & 0.19 & 0.58$\,\pm\,0.11$ & 74.2$\,\pm\,7.7$ & 0.21 & 0.73$\,\pm\,0.12$ \\
    $\lambda_{\rm gas}$< 0.051                &  \cellcolor{white}                                                     &    64.0$\,\pm\,5.5$ & 0.26 & 18.5$\,\pm\,4.9$ & 0.26 & 0.29$\,\pm\,0.08$ & 30.4$\,\pm\,6.6$ & 0.19 & 0.48$\,\pm\,0.11$ \\
    $\lambda_{\rm gas}$ > 0.051               &  \cellcolor{white}\multirow{-2}{*}{\ref{fig:slices:properties}.F}      &    101.3$\,\pm\,11.8$ & 0.17 & 52.0$\,\pm\,8.4$ & 0.19 & 0.51$\,\pm\,0.10$ & 70.8$\,\pm\,7.2$ & 0.22 & 0.70$\,\pm\,0.11$ \\
    
    \midrule
    
    $\beta_{\rm 3D}$ < 0.15        &  \cellcolor{white}                                                   &    54.8$\,\pm\,4.1$ & 0.21 & 22.8$\,\pm\,4.0$ & 0.26 & 0.42$\,\pm\,0.08$ & 35.7$\,\pm\,3.6$ & 0.23 & 0.65$\,\pm\,0.08$ \\
    $\beta_{\rm 3D}$ > 0.15        &  \cellcolor{white}\multirow{-2}{*}{\ref{fig:slices:dynamical-state}.A}    &   105.5$\,\pm\,8.5$ & 0.24 & 43.9$\,\pm\,10.3$ & 0.19 & 0.42$\,\pm\,0.10$ & 65.9$\,\pm\,9.5$ & 0.19 & 0.62$\,\pm\,0.10$ \\
    $f_{\rm sub}$ < 0.18             &  \cellcolor{white}                                                   &    72.0$\,\pm\,5.4$ & 0.20 & 25.2$\,\pm\,5.0$ & 0.25 & 0.35$\,\pm\,0.07$ & 52.2$\,\pm\,6.7$ & 0.19 & 0.72$\,\pm\,0.11$ \\
    $f_{\rm sub}$ > 0.18             &  \cellcolor{white}\multirow{-2}{*}{\ref{fig:slices:dynamical-state}.B}    &   89.5$\,\pm\,8.1$ & 0.25 & 40.5$\,\pm\,9.0$ & 0.20 & 0.45$\,\pm\,0.11$ & 49.3$\,\pm\,6.6$ & 0.24 & 0.55$\,\pm\,0.09$ \\
    $\cos\, \theta_{04}$ < 0.56               &  \cellcolor{white}                                                       &    64.1$\,\pm\,5.1$ & 0.23 & 16.3$\,\pm\,6.0$ & 0.16 & 0.25$\,\pm\,0.10$ & 26.7$\,\pm\,5.3$ & 0.28 & 0.42$\,\pm\,0.09$ \\
    $\cos\, \theta_{04}$ > 0.56               &  \cellcolor{white}\multirow{-2}{*}{\ref{fig:slices:dynamical-state}.C}        &   96.9$\,\pm\,9.9$ & 0.19 & 79.4$\,\pm\,8.8$ & 0.19 & 0.82$\,\pm\,0.12$ & 77.2$\,\pm\,7.7$ & 0.19 & 0.80$\,\pm\,0.11$ \\
    
    $\alpha_{\rm 3D}$ < 0.21        &  \cellcolor{white}                                                           &    58.6$\,\pm\,4.0$ & 0.22 & 23.9$\,\pm\,4.0$ & 0.25 & 0.41$\,\pm\,0.07$ & 41.2$\,\pm\,4.6$ & 0.20 & 0.70$\,\pm\,0.09$ \\
    $\alpha_{\rm 3D}$ > 0.21        &  \cellcolor{white}\multirow{-2}{*}{\ref{fig:slices:dynamical-state}.D}       &    101.9$\,\pm\,11.5$ & 0.17 & 40.6$\,\pm\,9.7$ & 0.20 & 0.40$\,\pm\,0.11$ & 60.3$\,\pm\,10.9$ & 0.13 & 0.59$\,\pm\,0.13$ \\
    $\alpha_{\rm 2D}\,(r= 0.2\,r_{500})$ < 0.047         &  \cellcolor{white}                                      &    58.9$\,\pm\,4.1$ & 0.18 & 21.1$\,\pm\,3.2$ & 0.28 & 0.36$\,\pm\,0.06$ & 34.4$\,\pm\,3.4$ & 0.23 & 0.58$\,\pm\,0.07$ \\
    $\alpha_{\rm 2D}\,(r= 0.2\,r_{500})$ > 0.047         &  \cellcolor{white}\multirow{-2}{*}{\ref{fig:slices:dynamical-state}.E}      &    103.8$\,\pm\,8.5$ & 0.25 & 45.3$\,\pm\,10.0$ & 0.20 & 0.44$\,\pm\,0.10$ & 65.2$\,\pm\,9.4$ & 0.20 & 0.63$\,\pm\,0.10$ \\
    $\alpha_{\rm 2D}\,(r=r_{500})$ < 0.14        &  \cellcolor{white}                                                      &    44.8$\,\pm\,2.7$ & 0.20 & 20.0$\,\pm\,2.5$ & 0.25 & 0.45$\,\pm\,0.06$ & 30.9$\,\pm\,3.4$ & 0.20 & 0.69$\,\pm\,0.09$ \\
    $\alpha_{\rm 2D}\,(r=r_{500})$ > 0.14        &  \cellcolor{white}\multirow{-2}{*}{\ref{fig:slices:dynamical-state}.F}        &    116.1$\,\pm\,8.8$ & 0.24 & 46.4$\,\pm\,10.8$ & 0.19 & 0.40$\,\pm\,0.10$ & 69.5$\,\pm\,11.3$ & 0.13 & 0.60$\,\pm\,0.11$ \\
    
    \midrule
    $z=0$ (75 clusters)              &  \cellcolor{white}                                                    &    18.0$\,\pm\,1.4$ & 0.23 & 7.5$\,\pm\,1.4$ & 0.25 & 0.42$\,\pm\,0.09$ & 12.9$\,\pm\,1.3$ & 0.27 & 0.71$\,\pm\,0.09$ \\
    $z=1$ (70 clusters)               &  \cellcolor{white}\multirow{-2}{*}{\ref{fig:slices:redshift}}        &    25.7$\,\pm\,2.9$ & 0.31 & 11.6$\,\pm\,3.5$ & 0.34 & 0.45$\,\pm\,0.15$ & 18.4$\,\pm\,3.2$ & 0.24 & 0.72$\,\pm\,0.15$ \\
    \bottomrule
    \end{tabular}
\end{table*}

\subsection{Complete sample}
\label{sec:results:complete-sample}

\begin{figure}
	\includegraphics[width=\columnwidth]{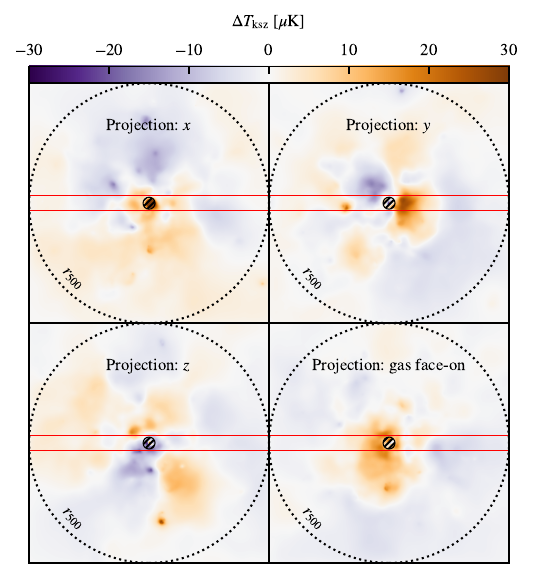}
    \caption{Maps showing $\Delta T_{\rm rkSZ}$ after stacking (and averaging) all MACSIS clusters considered in this work without any de-rotation (projections $x, y, z$ indicated at the top of each panel) and with ${\bf j}_0$ aligned and directed into the plane of the image (face-on projection). In all panels, we indicate $r_{500}$ as a black circle and with red rectangles the regions of the maps considered for computing the equatorial rotation profiles, spanning over $\pm\,r_{500}$ from the centre of the map in each direction.}
    \label{fig:rksz_gas_null}
\end{figure}

\begin{figure*}
	\includegraphics[width=2\columnwidth]{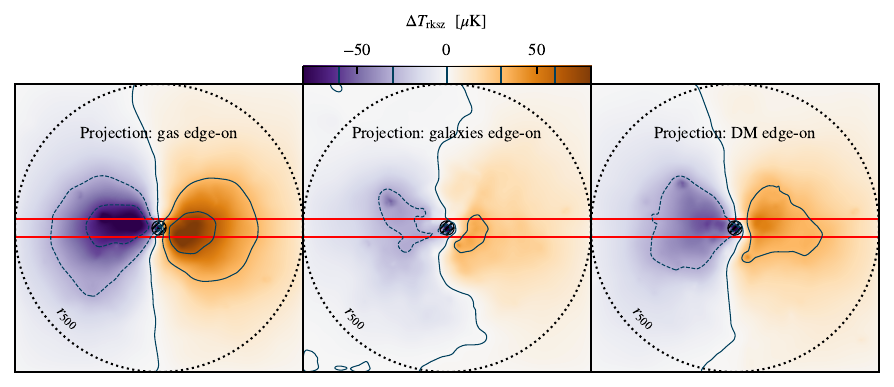}
    \caption{As in Fig. \ref{fig:rksz_gas_null}, but showing the stacked signal after de-rotating the clusters to an edge-on configuration. From left to right, the panels show the stacked signal when the alignment uses the angular momentum of the gas, the galaxies and the dark matter in $r_{500}$. The orange contours are evaluated for the same levels as in Fig. \ref{fig:rksz_model}. The largest rkSZ signal is obtained when orienting the clusters by aligning the gas spins.}
    \label{fig:rksz_gas_edge}
\end{figure*}

\begin{figure}
	\includegraphics[width=\columnwidth]{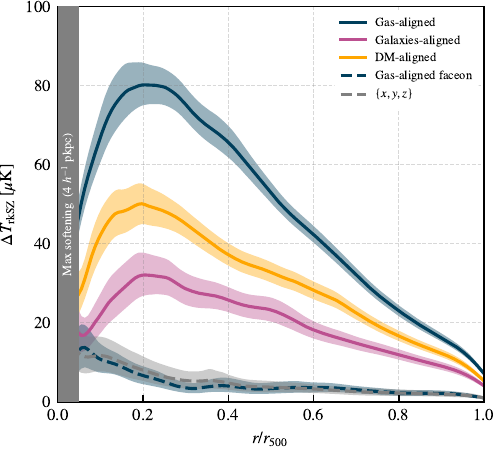}
    \caption{Rotational kSZ profiles obtained from the equatorial slices from the projections in Figs. \ref{fig:rksz_gas_null} (dashed lines) and \ref{fig:rksz_gas_edge} (solid lines) and folded about the centre of the map. The profiles are constructed from gas-aligned spins (blue), galaxies-aligned spins (orange) and dark matter-aligned spins (pink). We emphasize the gas-aligned face-on profile in blue, to be compared to its edge-on counterpart. The projections along the axes of the parent box are shown in grey. The grey vertical band represents the $r<0.05\, r_{500}$ region, which we exclude in our analysis. For each profile, the shaded bands represent the confidence intervals between the first and third quartiles, computed from $10^4$ realisations generated by bootstrapping the sample with repetition.}
    \label{fig:slices_projections}
\end{figure}

In Fig. \ref{fig:rksz_gas_null}, we show the maps for the mean rkSZ signal obtained from stacking the entire MACSIS sample (with the $M_{200}$ and $f_{\rm gas}$ selection of Section \ref{sec:datasets}) for the $\{x,y,z\}$ and the gas-face-on projections. Without de-rotation, or when the de-rotated angular momenta are oriented along the LoS (face-on), we expect the stacking method to suppress the rkSZ signal. The amplitude of the residual fluctuations in Fig.~\ref{fig:rksz_gas_null} is {$A_{\rm residual} \sim \langle \Delta T_{\rm rkSZ}^2 \rangle^{1/2} \approx 15~\mu$K}. Conversely, when the MACSIS clusters are de-rotated to an edge-on configuration, the rkSZ signal is stacked coherently and to produce amplitudes up to $A_{\rm max}\approx 80\,\mu$K, as shown in Fig. \ref{fig:rksz_gas_edge}. These maps were produced from an edge-on configuration aligning the angular momenta of the gas (left), galaxies (centre) and dark matter (left).

{Even without de-rotation, the maps in Fig. \ref{fig:rksz_gas_null} show features which could be mistaken for coherent motion. Their amplitude is often larger than that predicted by Poisson statistics, $A_{\rm residual, P} \approx 4~\mu$K}\footnote{{Assuming that the amplitude of residual fluctuation follows a Poisson distribution, then the variance should scale as $\propto 1/N$, where $N$ is the number of clusters in the sample. As more randomly-oriented clusters are stacked, we expect the amplitude of the fluctuations to decrease as $A_{\rm residual, P} \sim A_{\rm max} / \sqrt{N} \approx 4\, \mu$K, with $A_{\rm max}=80\,\mu$K and $N=377$.}}. {This result suggests that the amplitude of fluctuations are not generally Poisson-distributed. When probing the distribution of $A_{\rm residual}$, we found that the strongest signals in the maps of in Fig. \ref{fig:rksz_gas_null} originate from a subset of clusters where substructures produce intense rkSZ signals.} 
{We tested this hypothesis with two methods. (i) Firstly, we computed the \textit{median} of the $\{x,y,z\}$ and face-on projected signal pixel-wise. The median value is not affected by extreme data samples, unlike the average value. Using the median maps, we obtained fluctuations with typical amplitudes below $5~\mu$K, consistently with the Poisson estimate. (ii) In our second approach, we constructed a distribution of the peak amplitude, $\max(A_{\rm residual})=\max(|\Delta T_{\rm rkSZ}|)$, of the cluster maps and removed the clusters with high $\max(A_{\rm residual})$, above the 75$^{\rm th}$ percentile. The maps of the remaining clusters were stacked by averaging, and we recovered fluctuations of $5~\mu$K. Finally, we checked that the extreme objects that have been discarded in this test produce fluctuations $>15~\mu$K upon stacking. Such fluctuations are associated with substructures rich in dense gas and/or with extreme differential velocities along the LoS [see Eq. \eqref{eq:kSZ_definition}]. We leave an analysis of the impact of substructures on the rkSZ signal to future work.}

While the spatial distribution of the rkSZ amplitude is best probed using the maps in Figs. \ref{fig:rksz_gas_null} and \ref{fig:rksz_gas_edge}, the radial rkSZ profiles provide a quantitative comparison of the relative rkSZ amplitudes. In Fig.~\ref{fig:slices_projections}, we obtain the maximum amplitude (solid lines) when aligning the spin of the gas, $A_{\rm max}^{\rm (gas)}=80.2\,\mu$K (blue), while aligning to the spin of the galaxies produces an amplitude $A_{\rm max}^{\rm (galaxies)}=32.1\,\mu$K (pink), which is only $\simeq 40\%$ of the theoretical maximum signal. Aligning the spins of the dark matter halos produces an intermediate amplitude $A_{\rm max}^{\rm (DM)}=50.1\,\mu$K, {reproducing $\simeq 63\%$ of the maximal signal}. We summarise the measured rkSZ amplitudes $A_{\rm max}$ and the radius $r_{\rm max}$ where they peak in Table~\ref{tab:max_measured}. Although $A_{\rm max}$ varies significantly, in all cases we find $r_{\rm max}\simeq 0.2 \,r_{500}$, {which is in good agreement with \citetalias{CC02}.} The profile amplitudes of the edge-on configurations with all clusters can be clearly distinguished from the face-on and the $\{x,y,z\}$ set-ups, {which never exceed $\approx 15~\mu$K, even when objects with extreme fluctuations are included}. 

Given the results from stacking the MACSIS cluster sample \textit{in toto}, we now split the sample based on the the value of the cluster properties in Fig. \ref{fig:corner-plot-basic-properties}. For each property, we compute the median value and we define a high- and low-value sample. The rkSZ maps of clusters in are stacked separately for each of these two samples, allowing to probe difference in radial rkSZ profiles directly. These selection criteria are classified in three groups: the cluster properties at $z=0$ are presented in Section \ref{sec:results:cluster-properties}, the dynamical state metrics in Section \ref{sec:results:dynamical-state} and the redshift dependence in Section \ref{sec:results:z-dependence}. The results are also summarised in Table~\ref{tab:max_measured}. We conclude this discussion with a study of the rkSZ amplitude arising from differential motions in the ICM (Section \ref{sec:results:differential-motions}).

\subsection{Selection by cluster property}
\label{sec:results:cluster-properties}

\begin{figure*}
	\includegraphics[width=2\columnwidth]{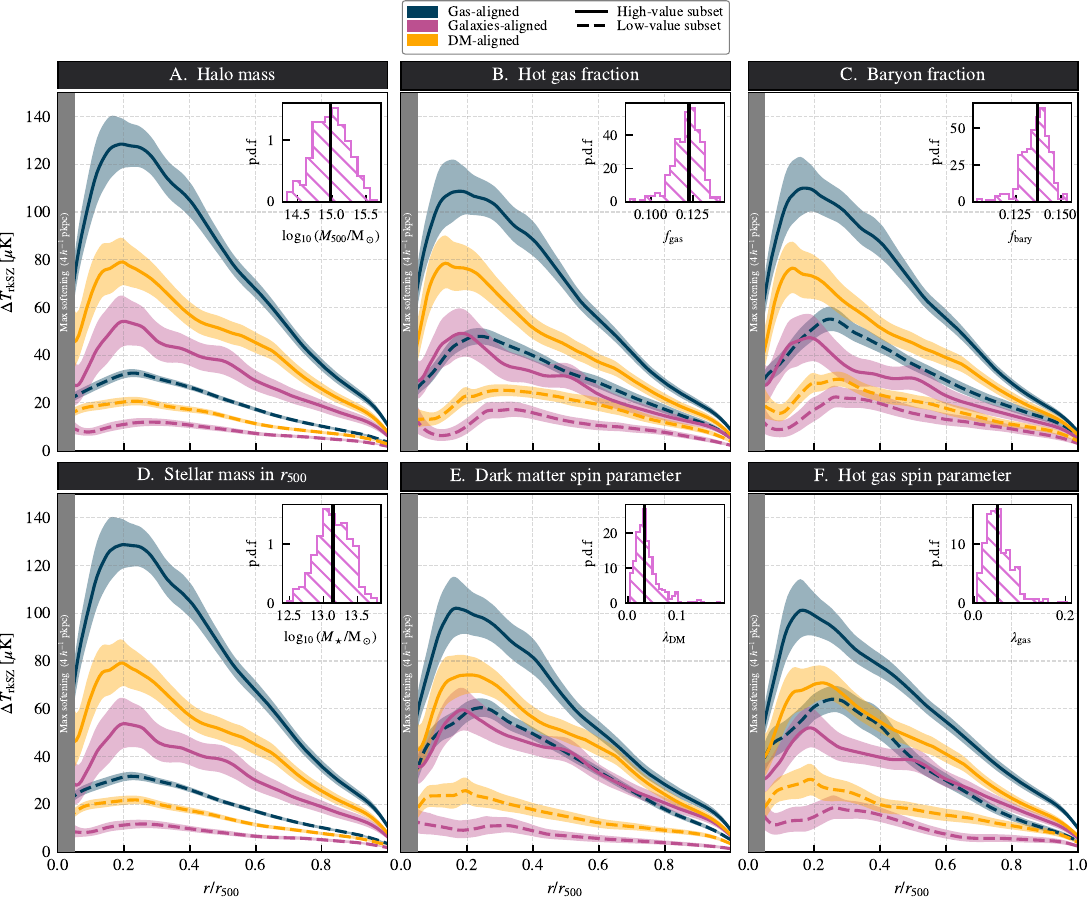}
    \caption{Radial rkSZ profiles in edge-on configuration for the MACSIS sample split according to the basic cluster properties of Section \ref{sec:alignment:properties}. The alignment methods are colour-coded as in Fig. \ref{fig:slices_projections}. The MACSIS sample is split into a high-value subset (solid lines), and a low-value subset (dashed lines). The rkSZ profiles are computed from stacking de-rotated rkSZ maps from clusters in each subset separately. The inset plot on the top-right of each panel shows the normalised p.d.f. of the quantity and the median, shown as a vertical solid line, defines the mass cut. At the top of each panel, we specify the property being examined, indexed from A to F.}
    \label{fig:slices:properties}
\end{figure*}

Based on the median $M_{500}$ of the MACSIS clusters, $9.7 \times 10^{14}~{\rm M}_\odot$, we split the sample into a high- and low-mass sample, and we produce the radial rkSZ profiles in Fig. \ref{fig:slices:properties}.A. Here, the solid lines refer to the stacked (averaged) high-mass sample, while the dashed lines indicate the low-mass samples. The colours represent the gas, galaxies and dark matter spin alignment methods, similarly to Fig. \ref{fig:slices_projections}, and are used consistently throughout this work. We found that the high-mass MACSIS clusters produce a rotational signal $\approx 4$ times stronger than the low-mass ones. This ratio remains the same regardless of the spin alignment method. Cluster-sized dark matter halos form hierarchically by accreting smaller objects, which introduce angular momentum into the system until the turnaround point. High-mass halos virialise later than low-mass ones and they experience a longer angular momentum growth phase \citep[e.g.][]{2002MNRAS.332..325P}. Assuming that the gas distribution traces the dark-matter potential, \citetalias{CM02} predicted a direct correlation between halo mass and rkSZ signal strength and our results in Fig. \ref{fig:slices:properties} confirm this. Moreover, this increase exceeds the prediction of the scaling relation in Eq.~\eqref{eq:kSZ-scaling}. Taking the first and third quartiles of $M_{500}$ as the centroids of the mass bins, the self-similar relation predicts $y_{\rm kSZ}$ to be $\approx 1.8$ times higher in the high-mass sample than in the low-mass sample. Although Eq.~\eqref{eq:kSZ-scaling} underpredicts the kSZ signal from cluster rotation in MACSIS by a factor of 2, we have shown that the mass-dependent trend is consistent with self-similar expectations, noting that these should only be used as guidelines. Self-similar scaling does not account for the dynamical state of the cluster. In Section \ref{sec:results:dynamical-state}, we will show that unrelaxed clusters produce a stronger signal and, since these objects also tend to be the most massive (see Fig.~\ref{fig:corner-plot-dynamical-state}), the rkSZ amplitude is likely to exceed the self-similar prediction.

The galaxies-aligned and the DM-aligned profiles have amplitudes 60 and 35\% lower than the gas-aligned profiles. In fact, we expect the gas-aligned profiles to produce the strongest possible signal, since the rotation of the ICM is made coherent by construction before stacking. Remarkably, choosing the angular momentum of the galaxies as proxy for the rotation axis of the gas suppresses $\approx$60\% of the signal to 54 $\mu$K for the high-mass sample. 
The MACSIS clusters in the low-mass bin therefore predict an rkSZ amplitude which is $\approx 10$ times larger than the estimate by \citetalias{CC02}. Our results are compatible with the amplitude found by \cite{2017MNRAS.465.2584B} from the MUSIC clusters of comparable mass and also agree well with the analytic estimates of \citetalias{CM02} for recent mergers.
{We discuss possible causes of the difference with \citetalias{CC02} in Section \ref{sec:summary}}, but we anticipate that {one of the main effects is the larger spin of the MACSIS clusters.} 

We then split the MACSIS sample into subsets with high and low hot gas fractions and show the radial rkSZ profiles in Fig. \ref{fig:slices:properties}.B. The clusters with $f_{\rm gas}$ above the median value of 0.12 produce a rotational signal 2.5 to 3 times stronger than clusters with low gas fractions. Since $y_{\rm rkSZ} \propto n_e$ (see Eq.~\ref{eq:kSZ_definition}), clusters with large hot gas content are expected to produce large $\Delta T_{\rm rkSZ}$ contributions to the rotational profiles. Moreover, the $f_{\rm gas}$-$M_{500}$ scaling relation, shown in Fig. \ref{fig:corner-plot-basic-properties}, suggests that clusters with high $f_{\rm gas}$ are also massive. Therefore, we expect $A_{\rm max}$ to increase with $f_{\rm gas}$ also because $f_{\rm gas}$ increases with $M_{500}$. The baryon fraction $f_{\rm bary}$ also affects the rkSZ amplitude in a similar fashion.

The following two results in our analysis consider the spin parameter of the hot gas and the dark matter. The spin parameter, defined in Eq.~\eqref{eq:spin-parameter}, probes the fraction of the kinetic energy in a system which is associated to rotational motion, as opposed to unordered dynamics, or turbulence in the case of the hot gas. We begin by splitting the MACSIS sample using the median value of $\lambda_{\rm DM}$ and we present the radial rkSZ profiles in Fig. \ref{fig:slices:properties}. For MACSIS, we find that fast dark matter rotators 
{($\lambda_{\rm DM} > 0.03$)} 
produce a stronger rkSZ signal than clusters with low $\lambda_{\rm DM}$, as expected in a scenario where the gas traces the DM during the angular momentum growth phase. 

Conversely, in Fig. \ref{fig:slices:properties} we show a larger kSZ contribution from rotational motion in cluster with high $\lambda_{\rm gas}$, and a weaker signal in clusters with $\lambda_{\rm gas} < 0.051$. When a large fraction of the hot gas kinetic energy is associated with unordered motion (thermalised ICM), such as in the low $\lambda_{\rm gas}$ sample, the rkSZ amplitude is smaller than in the cluster sample with a larger fraction of energy associated to coherent rotation.

\subsection{Selection by dynamical state}
\label{sec:results:dynamical-state}

\begin{figure*}
	\includegraphics[width=2\columnwidth]{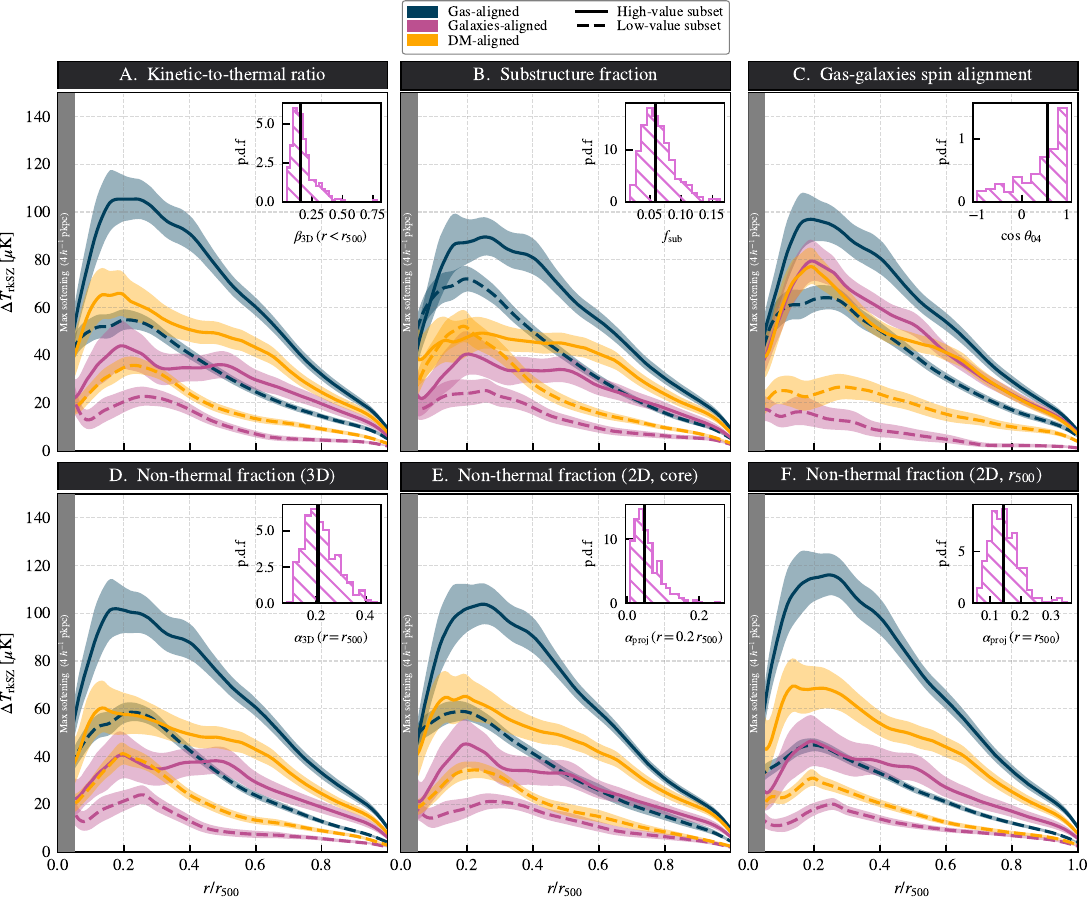}
    \caption{As in Fig. \ref{fig:slices:properties}, but splitting the MACSIS sample based on dynamical state descriptors (see Fig. \ref{fig:corner-plot-dynamical-state}).}
    \label{fig:slices:dynamical-state}
\end{figure*}

Following the results on the hot gas and dark matter spin parameters, we investigate the role of the substructures in the rotational kSZ signal. As in the previous results, we show in Fig. \ref{fig:slices:dynamical-state} the radial rkSZ profiles for clusters with substructure fraction below and above the median. Clusters with high substructure fractions tend to produce, on average, an rkSZ signal larger than clusters with a small substructure population.

The kinetic-to-thermal ratio measures the relaxation state of the hot gas in clusters and here we use it to define a thermodynamically relaxed sample, with $\beta_{\rm 3D} < 0.15$, and a non-relaxed sample, $\beta_{\rm 3D} > 0.15$, as shown in Fig. \ref{fig:slices:dynamical-state}.A. We find that non-relaxed clusters produce an rkSZ signal twice as strong as the relaxed. Therefore, the rkSZ amplitude is enhanced in clusters with active mergers, where the transfer of angular momentum drives the formation of dipole-like feature in the $\Delta T_{\rm kSZ}$ map. {The rkSZ amplitude is thus a proxy for the dynamical state of the cluster, as anticipated previously.}

We also split the MACSIS sample based on the angle between the angular momenta of the hot gas and the galaxies in $r_{500}$. Here, the value of $\cos \theta_{04}$ is used to define a gas-galaxies \textit{aligned} samples and a \textit{misaligned} sample. Focusing on the aligned sample, composed of clusters with $\cos \theta_{04} > 0.56$, we report an 18\% reduction in amplitude when de-rotating the maps using the galaxies as proxy for the rotation axis instead of using the gas angular momenta. For the misaligned sample ($\cos \theta_{04} < 0.56$), on the other hand, the amplitude is suppressed by 75\%. {This clearly highlights the importance of understanding the degree of alignment of gas and galaxy spins.}

To quantify the role of substructures in the defining the direction of the gas spin, we computed the angle between the galaxies spin and the hot gas in $r_{500}$ with and without substructures. We find that the values of $\cos \theta_{04}$ for the two scenarios have a median difference of 0.4 \% and $\approx 4 \%$ at the 90$^{\rm th}$ percentile. Given these results, we predict that the hot gas substructures in $r_{500}$ do not affect the orientation of the total angular momentum significantly, for most MACSIS clusters. We therefore predict that the substructure contribution to the de-rotation procedure is small enough to leave the rkSZ profiles largely unaltered and that nearly all the angular momentum is associated with the ICM.

Finally, we quantify the effect of non-thermal pressure on the rkSZ profiles by splitting the sample based on values of $\alpha$ from 3D and projected profiles as discussed in previous sections. The bottom row in Fig. \ref{fig:slices:dynamical-state} shows that high non-thermal pressure, possibly associated with turbulence, enhances the rkSZ amplitude by $\approx 40 \%$, compared to a cluster sample with low values of $\alpha$. When comparing the rkSZ amplitudes between high- and low-$\alpha_{\rm proj}$ samples, we find that the $\alpha_{\rm proj}(r=r_{500})$ selection (Fig. \ref{fig:slices:dynamical-state}.F) gives a larger difference, 71.3~$\mu$K, than the  $\alpha_{\rm proj}(r=0.2\, r_{500})$ selection (Fig. \ref{fig:slices:dynamical-state}.E), 44.9 $\mu$K. These values are computed for the gas-aligned configuration, and the same relative differences are found for the galaxies-aligned and DM-aligned scenarios.

\subsection{Redshift dependence}
\label{sec:results:z-dependence}

From simple self-similar scalings, the rotational kSZ amplitude increases with halo mass and redshift (see Eq.~\ref{eq:kSZ-scaling}). However, typical halo masses are not independent of redshift: clusters at low redshift  are on average more massive, since these have undergone cosmological accretion for longer (in a $\Lambda$CDM universe, this result can be interpreted as the halo-mass function shifting towards higher masses as structures-form hierarchically). Therefore, if we were to track a fixed population of clusters through redshift, the difference in rkSZ amplitude that would be measured would not just be due to a redshift dependence {of the dynamical quantities}, but a \textit{combined redshift and halo-mass} dependence.

\begin{figure*}
	\includegraphics[width=2\columnwidth]{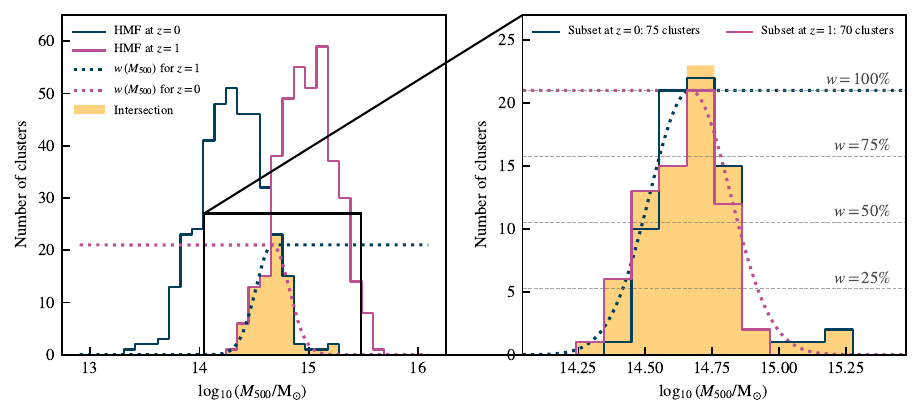}
    \caption{\textit{Left.} Halo mass functions for the complete MACSIS sample at $z=0$ (blue) and $z=1$ (pink). The weighting function $w$ is shown as dotted lines and its maximum value ($w=1$) is set to be the same as the peak of the Gaussian fit to the intersection region $N_\cap$ (yellow). \textit{Right.} Zoom-in view of the intersection region. Here, the halo mass functions are limited to objects that have been selected by the HMF-matching algorithm. We also show the quartile levels of the $w$ selection function as horizontal dashed lines. The median $M_{500}$ is $4.52 \times 10^{14}$ M$_\odot$ for the selected $z=0$ subset and $4.71 \times 10^{14}$ M$_\odot$ for the $z=1$ subset.}
    \label{fig:redshift_hmf_match}
\end{figure*}

\begin{figure*}
	\includegraphics[width=2\columnwidth]{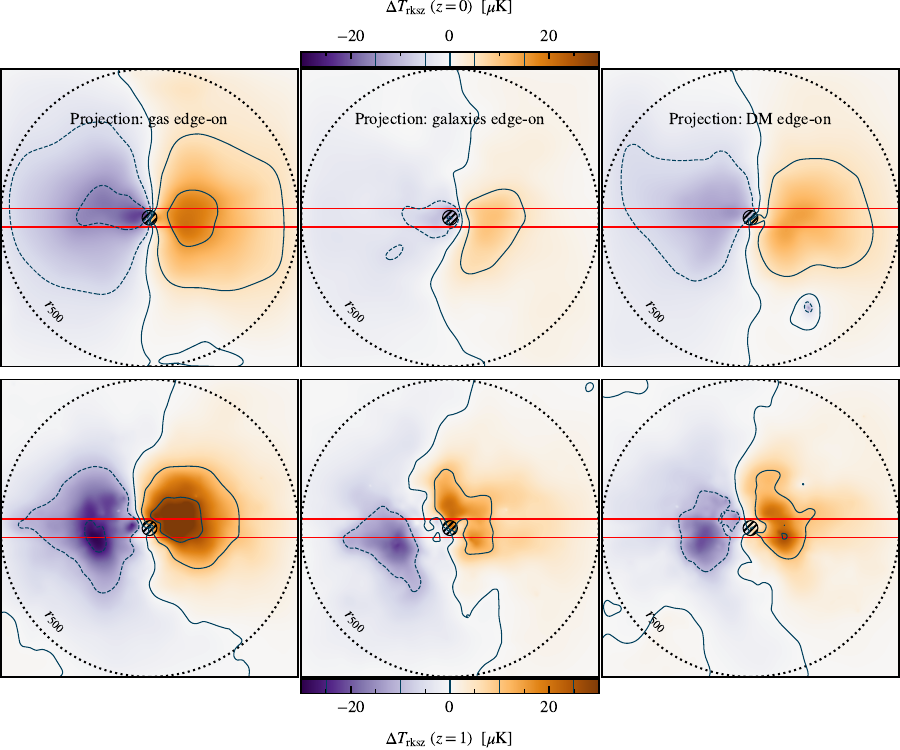}
    \caption{As in Fig. \ref{fig:rksz_gas_edge}, but comparing the edge-on configurations for the $z=0$ subset (75 clusters, top row) and the $z=1$ subset (70 clusters, bottom row). In all maps, the color bar limits are set to [-30, +30] $\mu$K and kept constant to allow a direct comparison of the rkSZ signal strength. We plot different contour levels to highlight the dipolar pattern at both redshifts: $\{0, \pm 5, \pm 15\}~\mu$K for $z=0$ and $\{0, \pm 10, \pm 25\}~\mu$K for $z=1$. The maximum rkSZ amplitude for the $z=1$ sample aligned with the gas spin ($\approx 30\, \mu$K) is larger than that of the $z=0$ sample ($\approx 20\, \mu$K).}
    \label{fig:maps:redshift}
\end{figure*}

We describe a method for matching the halo-mass function (HMF) with the aim to sample clusters of similar $M_{500}$ at different redshifts and probe the rkSZ effect at fixed halo mass. In this study, we consider the MACSIS sample at two redshifts, $z=0$ and $z=1$, and their halo mass functions shown in Fig. \ref{fig:redshift_hmf_match}. The HMFs for the MACSIS sample are not monotonically decreasing with halo mass as in e.g. \cite{2008ApJ...688..709T}, because the object selection is only mass-limited for the 90 most massive halos \citep[see][for further details]{2017MNRAS.465.2584B}. As a result, the HMFs at the two redshifts overlap only for a limited range of masses ($10^{14.25}<M_{500}/{\rm M}_\odot < 10^{15.25}$) and we highlight this intersection in yellow in Fig. \ref{fig:redshift_hmf_match} in the left panel. This intersection defines the mass range of our HMF-matched subset and the object count in each overlapping bin determines how many objects in the original mass functions must be selected. Given the halo mass function at the two redshifts $N(M_{500}, z=0)$ and $N(M_{500}, z=1)$, we define the intersection as
\begin{equation}
    N_{\cap}=\min\left[N(M_{500}, z=0), N(M_{500}, z=1)\right].
\end{equation}
We then fit a Gaussian profile $w_{\rm G}$ to $N_{\cap}$, normalise $w_{\rm G}$ to range between $[0, 1]$, and constructed weight functions $w(M_{500},z)$ as follows. For the $z=1$ weight function (blue dotted line), we combine piece-wise the half-profile with positive gradient up to its maximum ($M_{500, \rm peak}$) with a constant value of one elsewhere:
\begin{equation}
 w(M_{500},z=1) = \begin{cases} 
      w_{\rm G} & M_{500}\leq M_{500, \rm peak} \\
      1 & M_{500} > M_{500, \rm peak}.
   \end{cases}
\end{equation}
This functional form ensures that the selection includes all the high-mass clusters at $z=1$, but suppresses the sampling at low masses, which are not covered by the HMF at $z=0$. The weight function at $z=0$ (magenta dotted line) follows a similar prescription, but the piece-wise components are swapped. In this case, we achieve a mass-limited sampling for the low-mass halos at $z=0$, but we suppress the sampling at the high-mass end, which is not covered by the $z=1$ HMF. The functional form of the $z=0$ weight function is therefore
\begin{equation}
w(M_{500},z=0) = \begin{cases} 
      1 & M_{500} \leq M_{500, \rm peak} \\
      w_{\rm G} & M_{500} > M_{500, \rm peak}.
   \end{cases}
\end{equation}
In Fig. \ref{fig:redshift_hmf_match}, the right panel shows a zoomed-in view of the intersection histogram $N_{\cap}$ and the weight functions for the two redshifts. In the same plot, we also show four levels of the weight functions, where the HMF is sampled at 25, 50, 75 and 100\%, with grey dashed lines to guide the eye.

The following step includes combining the HMF with the weight function to obtain the \textit{number of objects} to be selected in each mass bin. We define this subset-HMF $N_{\rm subset}(M_{500}, z=0) = N(M_{500}, z=0)\times w(M_{500},z=0)$  and similarly for $z=1$. In mass bins where $w=1$, then we select all the objects with those masses, but for $w<1$ we choose $N_{\rm subset}(M_{500})$ objects at random from the $N(M_{500})$ in the original HMF. When downsampling the HMF, we choose a random selection to avoid bias towards high or low masses, depending on the redshift considered\footnote{We make the list of clusters for both redshift subsets publicly available online in our GitHub repository: \href{https://github.com/edoaltamura/macsis-cosmosim/redshift_samples}{github.com/edoaltamura/macsis-cosmosim/redshift\_samples} (see also the \textit{Data Availability} statement).}. 

The final selected subsets are shown in the right panel of Fig. \ref{fig:redshift_hmf_match} in the same colors as the HMF on the left. The $z=0$ subset contains 75 objects, located towards the low-end of the HMF at that redshift, while the $z=1$ subset contains 70 objects at the high-mass end of the HMF at that redshift. This method does not impose the subset size to match, and we therefore expect that the subset at the two redshifts may have a slightly different number of elements. The discrepancy is of the order of the Poisson noise, $\lfloor \sigma_{\rm P} \rfloor = 8$, and we find that imposing the same number of objects artificially does not change the final result for the rkSZ profiles. 

We note that a similar HMF-matching procedure was used by \cite{macsis_barnes_2017} and \cite{2020MNRAS.493.3274L} to combine the MACSIS sample at different redshifts consistently with the BAHAMAS data-set. Their method relies on a mass cut, which would be equivalent to $w(M_{500})$ being a step-function, where the jump-value is aligned with the minimum and maximum of the $N_{\cap}$ domain. This HMF-selection approach would overpopulate the high-redshift sample with low-mass objects, biasing the median $M_{500}$ towards lower masses, and overpopulate the low-redshift set with high-mass clusters, which would skew the p.d.f. in the opposite direction. While the differences between the HMF-matched and mass-cut methods are comparable to the Poisson noise due to the limited sample size of MACSIS, we emphasize that our HMF-matched approach is statistically robust and could deliver accurate mass-independent forecasts with much richer data sets from future large-volume hydrodynamic simulations.

This method introduced to match the HMF of the MACSIS sample is a novel procedure designed to produce a consistent mass coverage when comparing objects at different redshifts. The results are stable for HMF with a MACSIS-like shape. However, the functional form of $w(M, z)$ can be adapted to simulations with mass-limited HMFs, depending on the geometry of the overlapping region $N_{\cap}$.

\begin{figure}
	\includegraphics[width=\columnwidth]{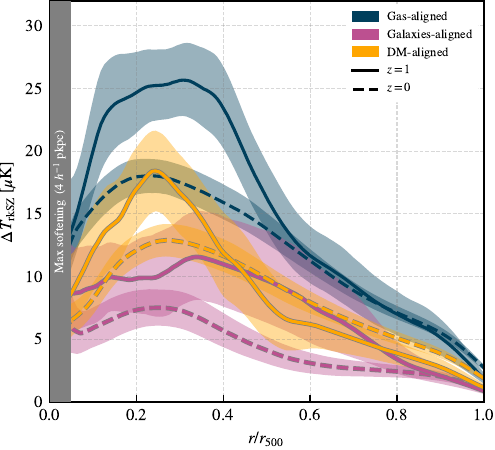}
    \caption{As in Fig. \ref{fig:slices:properties}, but comparing the HMF-matched sample at $z=0$ and $z=1$, derived from the rkSZ maps in Fig. \ref{fig:maps:redshift}. The galaxies- and DM-aligned profiles are shown with a grey edge for clarity.}
    \label{fig:slices:redshift}
\end{figure}

We show the rkSZ maps for the stacked cluster subset at $z=0$ and $z=1$ in Fig. \ref{fig:maps:redshift}; the corresponding equatorial profiles are reported in Fig. \ref{fig:slices:redshift}. The measured rkSZ amplitude at $z=1$ is found to be $\approx 1.4$ times larger than at $z=0$. In this instance, the self-similar scaling in Eq.~\eqref{eq:kSZ-scaling} overestimates this ratio, yielding $\left[E(z=1) / E(z=0)\right]^{5/3} = E(z=1)^{5/3} \approx 2.6$. Here, the mass dependence can be neglected because the HMF of the two subsets, and hence the median $M_{500}$, is matched by construction. By highlighting the redshift subsets in the corner plots in Figs. \ref{fig:corner-plot-basic-properties} and \ref{fig:corner-plot-dynamical-state}, we found that the $z=1$ have a larger median $\beta_{\rm 3D}$ value than the $z=0$ sample. In fact, most clusters at $z=1$ are captured during their accretion phase, when mergers cause them to be unrelaxed. This dynamical state leads to enhanced rkSZ signal, as we have shown in Section \ref{sec:results:dynamical-state} and panel A of Fig. \ref{fig:slices:dynamical-state}. The discrepancy in the amplitude between the MACSIS prediction and the self-similar scaling may be caused by the velocity scaling leading to Eq.~\eqref{eq:kSZ-scaling}, which expresses the circular velocity of virialised halos. In hydrodynamic simulations, this value is usually found to be higher than the tangential component of the velocities associated with the bulk rotation of the ICM \citep[see e.g.][]{2017MNRAS.465.2584B}, causing the self-similar scaling to have a stronger redshift dependence than we measured. We defer a detailed redshift study to future work.

While clusters similar to those in MACSIS at high redshift may contribute to the detected SZ signal more strongly, especially on small angular scales, their rkSZ amplitude is also greatly reduced when the de-rotation is based on the spin of the galaxies. Examining the galaxies-aligned profiles in Fig, \ref{fig:slices:redshift}, we measure amplitudes of $7.5~\mu$K for $z=0$, and $11.6~\mu$K for $z=1$. Both values are comparable to the rkSZ amplitude from the unordered substructure motions $\approx 5~\mu$K obtained from the same subsets ($\{x, y, z\}$-projections), and could potentially make the detection of the rkSZ signals from stellar proxies more challenging, as explained in Section \ref{sec:results:differential-motions}.

\subsection{Differential motions}
\label{sec:results:differential-motions}
The growth of galaxy clusters is driven by the infall of substructures, which contribute to creating a complex dynamic environment. When selecting a particular aperture to compute cluster properties, and particularly the angular momentum vector orientation, we actually obtain averaged quantities over the $r_{500}$ sphere. Clearly, the substructure clumps (individually) and their peculiar radial motion carry angular momentum which may be significantly different in magnitude and orientation than the aperture average. The \textit{differential motion} of ionised gas clouds inside the selected aperture can have an effect on how the angular momentum is estimated, and therefore on the de-rotation procedure too. Stacking de-rotated rkSZ maps heavily mitigates this effect, however, we have shown in Fig. \ref{fig:rksz_gas_null} that residual motions still appear in absence of de-rotation and even when the LoS is parallel to $\mathbf{J}$. The corresponding rkSZ profiles in Fig. \ref{fig:slices_projections} (grey and blue dashed lines) demonstrate that differential substructure motions can produce amplitudes as large as 20 $\mu$K near the centre. Observational studies relying on the angular momentum of galaxies to de-rotate the kSZ maps may often measure \textit{stacked} rkSZ amplitudes of $\approx 20~\mu$K (see profiles for galaxies-aligned, low-value subsets in Figs. \ref{fig:slices:properties} and \ref{fig:slices:dynamical-state}), which are comparable to the those produced by unordered differential motions. The question of whether the rkSZ features are due to ordered rotation or unordered substructure motions, even after stacking, makes the study of $\{x, y, z\}$-projection maps necessary and compelling.

\begin{figure*}
	\includegraphics[width=2\columnwidth]{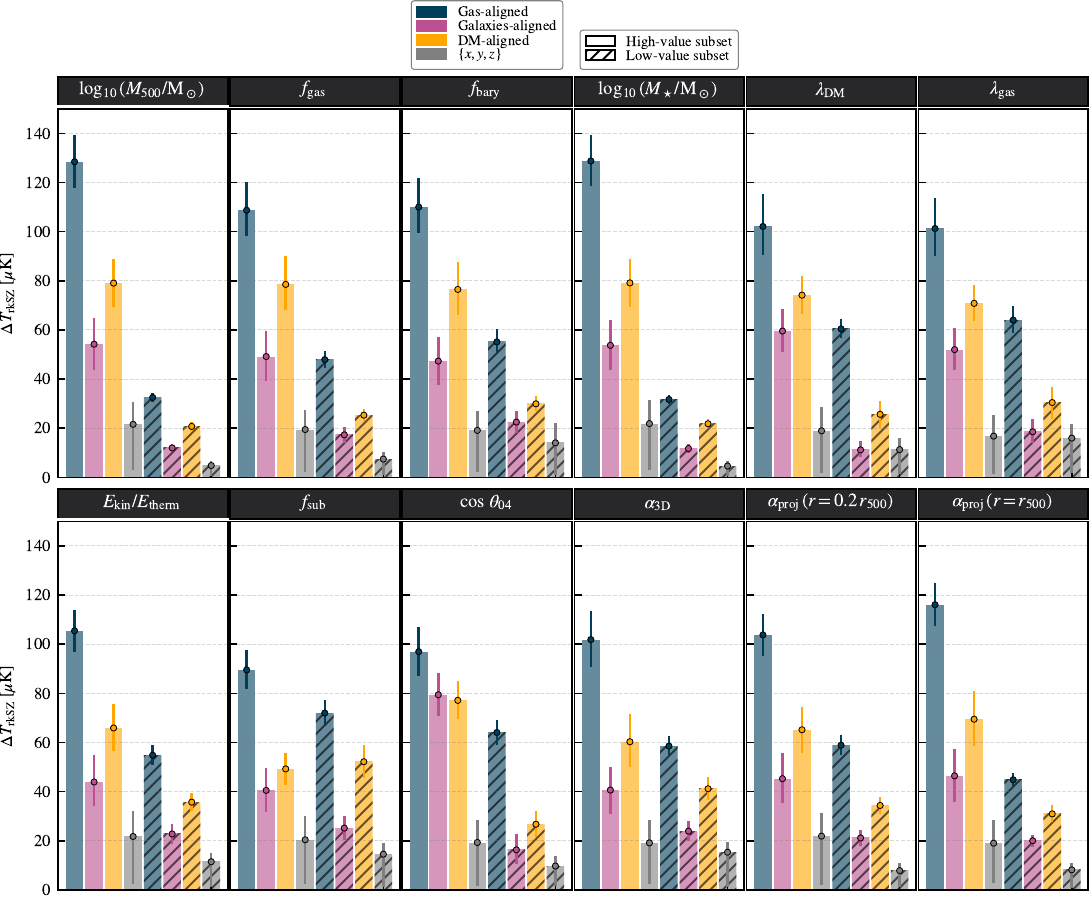}
    \caption{Summary of the rkSZ amplitude, $A_{\rm max}$, of the profiles in Figs. \ref{fig:slices:properties} (top row) and \ref{fig:slices:dynamical-state} (bottom row). The colours indicate the de-rotation configuration, as above; the high- and low-value subsets are represented by a solid colour and a hatch pattern respectively. In addition to the Gas, galaxies and DM alignment configurations, we show the amplitude of the non-de-rotated projections, quantifying the rkSZ signal due to differential motions for each population subset.}
    \label{fig:differential-motions}
\end{figure*}

Using the median profiles explored in this section, we record the maximum amplitude or the rkSZ signal and we compare it with that of the $\{x, y, z\}$-projection profiles. In Fig. \ref{fig:differential-motions}, we show these results for the cluster subsets next to the corresponding three non-de-rotated projections averaged together. Identifying the rotational signature with a high confidence level demands that its amplitude is greater than the signal due to spurious features originating from unordered motion in the cluster's rest frame. An example of high-amplitude ratio is obtained when splitting the cluster population by $M_{500}$, in the top-left panel of Fig. \ref{fig:differential-motions}. The signal from differential motions of 188 stacked high-mass clusters\footnote{The MACSIS sample, reduced to 377 clusters after the $M_{200}$ and $f_{\rm gas}$ cuts, is split by the \textit{median} of a cluster property, producing two subsets of 189 and 188 clusters.} has an amplitude $A_{\rm max}^{\{x,y,z\}} \approx 20\, \mu$K, while $A_{\rm max}^{\rm (gas)}=128\, \mu$K is $\approx 6$ times larger. Conversely, the low-$\lambda_{\rm DM}$ and low-$\lambda_{\rm gas}$ samples show $A_{\rm max}^{\{x,y,z\}} \approx A_{\rm max}^{\rm (galaxies)}$, suggesting that the rkSZ signal after stacking these objects, obtained with a galaxy-based de-rotation criterion, will likely be due to unordered motion rather than coherent rotation. This result may impact the significance of rkSZ measurements, highlighting once again the importance of using a robust de-rotation method to maximise the overall amplitude upon stacking.

Improving the accuracy of the de-rotation, however, might not be sufficient to achieve a high $A_{\rm max} / A_{\rm max}^{\{x,y,z\}}$ ratio where the signal is intrinsically low, e.g. in low-mass clusters. The effect of differential motions could, however, be reduced in two ways: (i) increasing the number of stacked clusters would suppress the random features and (ii) varying the aperture used in the calculation of the bulk velocity (and angular momentum) would shift the rest frame of the cluster (and the de-rotation alignment) in phase space, suppressing some features from unordered motion and enhancing others, allowing to estimate the underlying coherent rotational signature more reliably. We delegate the detailed investigation of the impact of sample size and aperture selection to future work.

Based on this argument, we comment on the evidence for cluster rotation measured by \cite{2019JCAP...06..001B}. The sample used in their study includes low-redshift ($z \in [0.02-0.1]$) and  $M_{500} \in [10^{14} - 10^{15}]$ M$_\odot$ clusters, compatible with our MACSIS low-mass subset de-rotated using galaxy spins. Based on results in Table~\ref{tab:max_measured}, we predict $A_{\rm max}^{\rm (galaxies)} = (32.5 \pm 1.7)\, \mu$K, which is comparable to their measured amplitude. For this subset, we estimate a differential motion amplitude of $A_{\rm max}^{\{x,y,z\}} \approx 5\, \mu$K, which is $\approx 6$ times lower than the rkSZ amplitude. Our prediction supports their $2\sigma$ detection claim, but we recommend caution when using their result, given the limited sample sizes of 6 and 13 objects.

\section{Fitting a analytic model}
\label{sec:results:fitting}

We now provide a analytic model which can represent the rkSZ profiles of massive clusters. To obtain these results, we follow the procedure used by \cite{2017MNRAS.465.2584B}, which includes two steps: (i) fitting a \citeauthor{2006ApJ...640..691V} model to the 3D number density profile $n_e(r)$ and (ii) fitting an angular velocity profile $\omega(r)$ to the projected (2D) rkSZ maps, assuming the $n_e(r)$ found previously.

\begin{figure*}
	\includegraphics[width=2\columnwidth]{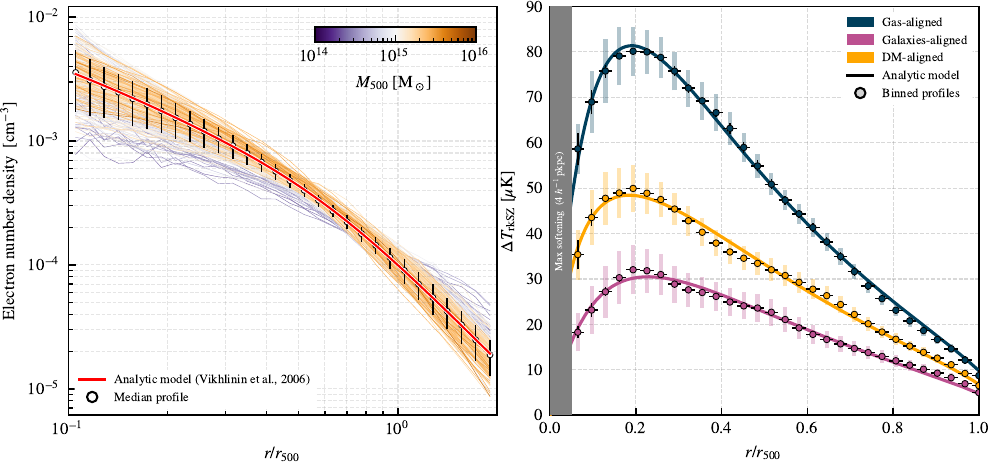}
    \caption{\textit{Left panel}. Density profiles (hot gas, $T>10^5$ K) for the MACSIS clusters colour-coded by $M_{500}$. The median profile is indicated by black markers and fitted by a \protect\citeauthor{2006ApJ...640..691V} model, shown in red. The black error bars indicate the first and third quartiles of the scaled density at each radial bin. \textit{Right panel.} The rkSZ profiles obtained after stacking all clusters (as in Fig. \ref{fig:slices_projections}), with colors indicating the de-rotation criterion as in previous figures. The profiles are down-sampled to 64 radial bins and fitted by the analytic rkSZ model in Eq.~\eqref{eq:kSZ-projection}. The black error bars indicate the 10$^{\rm th}$ and 90$^{\rm th}$ percentiles of the variation of the median profile in each bin, and the thicker coloured bars show the first and third quartiles of the profile from bootstrap resampling, as in Fig. \ref{fig:rksz_gas_edge}.}
    \label{fig:density-profiles}
\end{figure*}

\begin{enumerate}
    \item \textbf{Number density profiles}. As a first step, we divide the hot ($T>10^5$ K) gas particles in 30 radial shells with log-spaced radii ranging from $0.05\, r_{500} -\, r_{500}$. The mean density $\rho(r)$ in each shell is therefore given by the total mass divided by the volume of the shell. From the density profile, $n_e(r) = \rho(r) / (\mu \, m_{\rm H})$, where the $\mu=1.14$ is the mean atomic weight and $m_{\rm H}$ is the mass of the hydrogen atom. In our procedure, we do not directly compute $n_e(r)$ and we scale $\rho(r)$ by the density profiles to the critical density of the Universe, and the radius by $r_{500}$. Using this method, we obtain the dimensionless density profile for each MACSIS cluster in the selection subsets.

    {To reduce the degeneracy of the \citeauthor{2006ApJ...640..691V} model, we propose constraining the range of the parameters and fixing the slope $\alpha$. For the MACSIS clusters, we found that imposing the bounds reported in Table~\ref{tab:mass-dependence-fit-params} greatly stabilise the model (see Fig. \ref{fig:density-profile-mass-bins}).}
    
    \begin{table}
    \centering
    \caption{{Constraints imposed on the parameters in the \protect\citeauthor{2006ApJ...640..691V} model (columns 2-3), and linear-fit parameters ($a, b$) describing the dependence of the $\mathcal{V}_6$ parameters on halo mass (columns 4-5). In all our fits to electron number density profiles, we fix $\alpha=1.5$. $^\star$For the $z=1$ sample, we raise the upper bound of the cool-radius to $r_c = 0.28\, r_{500}$ to match the larger and shallower density core.}}
    \label{tab:mass-dependence-fit-params}
    \begin{tabular}{lcccc}
    \toprule
    \multicolumn{1}{c}{\bf Parameters}    &   
    \multicolumn{2}{c}{\bf Constraints} &
    \multicolumn{2}{c}{\bf Mass dependence} \\
    \cmidrule(rl){2-3} \cmidrule(rl){4-5}
    \multicolumn{1}{c}{$\mathcal{V}_6$} & Min & Max & $a$ & $b$ \\
    \midrule
    $n_0 / (10^{-3}\, {\rm cm^{-3}}$) & 0.9 & 9.0 & -6.47 & 103 \\
    $r_c /r_{500}$ & 0.01 & 0.18$^\star$ & 0.139 & -1.99\\
    $r_s /r_{500}$ & 0.50 & 0.75 & -0.0847 & 1.89\\
    $\alpha$ \quad (fixed) & 1.5 & 1.5 & 0 & 1.5\\
    $\beta$ & 0.3 & 0.6 & 0.115 & -1.29\\
    $\varepsilon$ & 2 & 3 & 0.275 & -1.45\\
    \bottomrule
    \end{tabular}
    \end{table}   
    
    
    {Having fixed $\alpha=1.5$, we fit a 5-parameter ($\mathcal{V}_{\rm 5}$)} \citeauthor{2006ApJ...640..691V} model to the median profile via the L-BFGS-B optimisation method \citep{zhu1997algorithm} implemented in \textsc{Scipy} \citep{virtanen2020scipy}. An example of this procedure is shown on the left panel of Fig. \ref{fig:density-profiles}, where we included all MACSIS clusters. The 6 best-fit parameters $\mathcal{V}_6$ for the cluster sample in Fig. \ref{fig:density-profiles} and those using other cluster selection criteria are summarised in Table~\ref{tab:vikhlinin_z0}.
    
    \item \textbf{Angular velocity profiles}. Given the priors $\mathcal{V}_6$ setting the density profile, which we keep fixed, we now proceed in the evaluation of the projection integral in Eq.~\eqref{eq:kSZ-projection} to find the best-fit parameters $\mathcal{B}_{\rm g}=\{v_{\rm t0}, r_0, \eta\}$ for $\omega(r)$. We now fit this template to the median rkSZ profiles in Figs. \ref{fig:slices_projections} to \ref{fig:slices:redshift} to obtain a analytic model for each subset sample of clusters we introduced. Unlike for the number density profile, here we use the Sequential Least Squares Programming \citep[SLSQP,][]{kraft1988software, nocedal2006numerical} method to optimise the fitting functional form, since we found a faster convergence time compared to the L-BFGS-B method.
    We found that the $\omega(r)$ model in Eq.~\eqref{eq:omega-profile} allows for great flexibility to match the rkSZ profiles well at all radii, due to the unconstrained $\eta$ parameter, as we demonstrate in the right panel of Fig. \ref{fig:density-profiles}. As in the previous steps, we report the best-fit parameters $\mathcal{B}_{\rm g}$ for the generalised angular velocity profile in Table \ref{tab:baldi_z0}.
\end{enumerate}

\section{The rkSZ temperature power spectrum}
\label{sec:power-spectrum}
The results of the previous sections deliver a simple prescription for the electron density and angular velocity profiles of clusters at varying redshifts. {In principle,} these can be directly used to predict the expected temperature power spectrum contribution using a halo model approach, improving on the treatment given in \citetalias{CC02}. {We outline and discuss the framework to carry out such a calculation, leaving a full calculation to future work with large-volume hydrodynamic simulations.}

The main starting point for the computation of the rkSZ temperature power spectrum calculation is the halo model \citep[e.g.,][]{Sheth1999,Seljak:2000gq, Cooray2002}, which allows to define the comoving number density of halos as a function of mass and redshift, ${\rm d} N(z, M)/{\rm d}M{\rm d}V$. The treatment then is essentially like for the standard $y$-distortion power spectrum \citep{Refregier2000, Hill2013}, but with the Legendre transform of the cluster $y$-profiles being replaced by the corresponding transforms of the rkSZ profile. For the 1-halo contribution to the CMB temperature power spectrum, this then reads \citepalias{CC02}
\begin{align}
C_{\ell}^{\rm 1h} 
&=\int_{0}^{z_{\rm max}} {\rm d}z \frac{{\rm d}V}{{\rm d}z}\int_{M_{\rm min}}^{M_{\rm max}} {\rm d}M \frac{{\rm d}N}{{\rm d}M{\rm d}V} \,\frac{2}{3}\, \left|\, y^{\rm rkSZ}_{\ell}(M,z)\, \right| ^2.
\label{eq:yy_total}
\end{align}
The factor of $2/3=\langle\sin^2 i\rangle$ arises from the average over a random distribution of inclination angles, $i$. This expression directly uses the fact that the rkSZ signal causes a simple temperature perturbation, $\Delta T/T$, on the sky, see Eq.~\eqref{eq:kSZ_definition}.
The Legendre transform of the rkSZ signal profile can be expressed as
\begin{subequations}
\label{eq:yy_transfor}
\begin{align}
\left|\, y^{\rm rkSZ}_{\ell}(M,z)\, \right| ^2
&\simeq\frac{\pi}{2}\,
\left[
\int_0^{\theta_{200}}
\eta(\theta)\,J_1(\ell\,\theta)\,{\rm d}\theta
\right]^2
\\[1mm]
\eta(\theta)&=
\frac{2\sigma_T}{c}R(\theta)\,\int_{R(\theta)}^{r_{200}} \frac{n_e(r)\,r\,\omega(r){\rm d}r}{\sqrt{r^2-R(\theta)^2}}
\end{align}
\end{subequations}
where we recast the expressions of \citetalias{CC02} {to match our formalism}. Here, $J_1(x)$ is the Bessel function of the first kind. We also use the angular diameter distance, $d_A(z)$, to obtain the radius $R(\theta)=d_A \theta$ via the angular scale $\theta$. For readability, we suppress the explicit dependency of the variables on mass and redshift.

In \citetalias{CC02}, the electron density profile was modeled using the hydrodynamic equilibrium assumption in a Navarro-Frenk-White \citep[NFW,][]{1996ApJ...462..563N} dark matter profile. In addition, the angular velocity profile was assumed to be given by simple solid body rotation {(i.e., $\omega \sim$ constant, see also Appendix \ref{app:profile-fits})}. Both aspects can be improved upon using the results {presented in this work}.

{To model the rotational kSZ signature using the halo model, we require both $n_e(r)$ and $\omega(r)$ as functions of $M_{500}$ and redshift. Within the present simulations, it is difficult to assess the detailed redshift dependence of the profiles, and therefore we recommend using the scalings obtained at $z=0$ for all redshifts. {In the future, large-volume ($\, \gtrsim 1$ Gpc$^3$) simulations will provide more accurate estimates of the rkSZ amplitude over a range of redshifts.}}

Consulting Table~\ref{tab:baldi_z0}, for the circular velocity profile in the gas-aligned case relevant here, we find that the parameters $r_0$ and $\eta$ only depend weakly on the cluster mass. We thus recommend the average values $\bar{r}_0\simeq 0.16\,r_{500}$ and $\bar{\eta}\simeq 2.0$ for the gas-aligned case. For $v_{\rm t 0}$, a mass-dependence is present, however, given the limited mass resolution of the simulation, we recommend using $\bar{v}_{\rm t 0}\simeq 1.5\,  v_{\rm circ}$. 
These simple scaling relations provide a reasonable representation of the simulation results, and broadly should give a signal that is about one order of magnitude larger than in \citetalias{CC02}.

{
For the electron density profile, from Table~\ref{tab:vikhlinin_z0} we find indications for a dependence of the profile parameters on cluster mass. To verify this dependence, we split the MACSIS sample in 8 logarithimcally spaced mass bins and then computed the median density profiles of the clusters in each bin. However, the density model parameters are strongly degenerate, motivating us to adopt $\alpha=1.5$ in all our fits. We show the fit model for each mass bin in Fig.~\ref{fig:density-profile-mass-bins}).}

{
We find that, for the mass range $\log_{10}(M_{500} /{\rm M}_\odot) \in [14.4, 15.6]$, the density fit parameters ($\mathcal{V}_i \in \mathcal{V}_6$) can be generally described by a linear function with gradient $a$ and intercept $b$, expressed as
\begin{equation}
    \label{eq:Ne_approximation}
    \mathcal{V}_i(M_{500}) = a \times \log_{10}(M_{500}/{\rm M}_\odot)  + b.
\end{equation}
The linear parameters $a$ and $b$ are reported in Table~\ref{tab:mass-dependence-fit-params} and we show the individual scaling relations in Fig. \ref{fig:density-profile-mass-bins-parameters}.
}

Together with the standard mass and redshift dependencies of $M_{500}$, $r_{500}$, $v_{\rm circ}$ and $d_A$, these relations can be used to compute $y^{\rm rkSZ}_{\ell}$ from Eq.~\eqref{eq:yy_transfor}. From Eq.~\eqref{eq:yy_total}, one can then obtain the auto-power spectra caused by the rkSZ effect, e.g., by using {\tt CLASS-SZ} \citep{Bolliet:2017lha} to compute the average over the mass function. We note that the relations in Eq.~\eqref{eq:Ne_approximation} are derived from a cluster sample with a very narrow mass range and they may not be valid for values of $M_{500}$ outside the MACSIS range. We therefore recommend calibrating the $\mathcal{V}_6$ parameters for mass-limited cluster data sets, spanning over a wider mass range.

{A detailed exploration of the dependence of the rkSZ power spectrum signal on various parameters will be left to future work. However, we comment of a few important aspects. Firstly, one does expect contributions from the 2-halo term to become relevant at large angular scales \citepalias[see also][]{CC02}. To compute this signal, a model for the halo bias with respect to linear theory must be provided. This treatment also must account for the effect of spin alignments, which can affect the total amplitude of the 2-halo contribution in non-trivial ways, requiring additional investigation.}
{Secondly, here we have only provided a prescription for the 1-halo contribution from the large-scale rotation. As our analysis has revealed, significant kSZ contributions also arise from internal motions of substructures (see Section~\ref{sec:results:differential-motions}). The 1-halo kSZ contribution will therefore be enhanced by these effects, in particular at small scales corresponding to $\ell \gtrsim {\rm few}\times 10^3$.}
{Finally, the rkSZ is expected to add an irreducible noise floor to studies of the moving-lens effect \citep{Birkinshaw1983, 2019PhRvL.123f1301H} and cosmological vorticity modes \citep{2023arXiv230111344C}, possibly biasing the inference. Given that the estimates of \citetalias{CC02} seem to be on the low end, extra effort should be made to improve the modeling of the rkSZ contribution for future applications.} {We leave a full calculation and discussion of the temperature power spectrum to future work.}

\section{Discussion and Conclusions}
\label{sec:summary}

Cluster rotation can be detected via its dipole-like kSZ signature, provided that the orientation of the angular momentum orientation of the gas is well estimated, or reconstructed. Our study of the rkSZ effect with MACSIS clusters complements that of \cite{2017MNRAS.465.2584B} where a sample of six low-mass relaxed galaxy clusters was selected from the MUSIC simulations. To achieve these results, we followed the methodology in Fig. \ref{fig:stacking_workflow}. Then, in Section \ref{sec:results}, we study the rkSZ profiles over a mass range $M_{500} \sim 10^{15-16}$ M$_\odot$, wider than that in \cite{2017MNRAS.465.2584B}, and showed how the rkSZ signal strength varies when clusters are selected by their global properties or dynamical state descriptors.

Our rkSZ amplitudes are comparable to the study from the MUSIC simulations \citep[$\approx 30-50~\mu$K,][]{2017MNRAS.465.2584B}, stressing that this value should be compared the low-mass (Fig.~\ref{fig:slices:properties}.A) and the relaxed (Fig.~\ref{fig:slices:dynamical-state}.A) subsets. Moreover, our high-mass subset shows an rkSZ amplitude close to the regime of a maximally rotating cluster merger, used by \citetalias{CM02} to set an upper bound ($\approx 200~\mu$K) to the typical kSZ contribution from bulk rotation. Overall, this study shows agreement with the rkSZ signal strength at the relaxed, low-mass end of the cluster population \citep{2017MNRAS.465.2584B} and at the unrelaxed, high-mass end (\citetalias{CM02}). However, we note that none of these works predicts an rkSZ amplitude as small as \citetalias{CC02}. From MACSIS, we measured a rkSZ amplitude 10 times larger than the \citetalias{CC02} estimate of $\approx 3~\mu$K, computed for a cluster with virial mass $5\times 10^{14}$ M$_\odot$ at $z=0.5$. Assuming their cosmological parameters and our self-similar relations, we can re-scale their value to match the halo mass and redshift of our sample. This calculation indicates that rkSZ amplitude at $z=0$ would be lower by 40\% at their cluster mass; up-scaling this value self-similarly in mass to match the median value for MACSIS ($M_{200}=1.4\times 10^{15}$ M$_\odot$) would increase the amplitude by a factor of 2, giving an overall signal of $\Delta T_{\rm CC02}\approx 3.6~\mu$K. We will compare this value with $80.2~\mu$K obtained from the \textit{all-clusters} sample (see Table~\ref{tab:max_measured}). We identified three parameters which may have been underestimated by \citetalias{CC02}. 
(i) The Universal \textit{baryon fraction} from their cosmology is 10\% lower than ours, causing the cluster gas mass to be underestimated by a factor of 0.9 \citep[see also][]{2000PhRvD..62j3506C}. Using the mass scaling in Eq.~\eqref{eq:kSZ-scaling}, the correction to the rkSZ amplitude is expected to be $\delta(f_{\rm b})\approx 1.1$.
(ii) they used the value of $\approx 36~{\rm km\, s^{-1}}$ for the \textit{tangential velocity} at $\approx 0.2$ Mpc, while we found $\approx 90~{\rm km\, s^{-1}}$  at $r/r_{500} = 1/5$, i.e. the radius of the maximum amplitude, which could increase the rkSZ signal strength linearly by a factor of $\delta(v_{\rm tan})\approx 2.5$. 
Finally, (iii) they assume the mean \textit{spin parameter} for dark-matter halos $\bar{\lambda}_{\rm DM} \approx 0.04$, while we recommend using the hot gas spin parameter instead, which has a median value of $\lambda_{\rm gas} \approx 0.05$, leading to an additional correction of $\delta(\lambda)\approx 1.2$. 
After combining these corrections, we find $\delta(f_{\rm b})\, \delta(v_{\rm tan})\, \delta(\lambda) \approx 3.3$, which is still not sufficient to explain the factor of 10 difference from our measurement. However, we have shown in Fig.~\ref{fig:slices:properties}.F that the rkSZ amplitude is dependent on $\lambda_{\rm gas}$, which can reach values of $\approx 0.15$. Clusters with high spin parameters are abundant: they could indeed contribute significantly to this estimate since they form an extended tail in the log-normal-like distribution at most halo masses \citep{2010MNRAS.404.1137B}. Now, assuming $\lambda_{\rm gas} \approx 0.15$, we find that \citetalias{CC02} may have underestimated the spin parameter by a factor of $\delta(\lambda)\approx 3.6$. When combined, these considerations lead to $\delta(f_{\rm b})\, \delta(v_{\rm tan})\, \delta(\lambda) \approx 10$, which could reconcile the prediction by \citetalias{CC02} with our study and \cite{2017MNRAS.465.2584B}.

We now highlight the following key findings from this work:
\begin{enumerate}
	\item \textbf{Mass dependence.} High-mass clusters produce a larger rkSZ amplitude than low-mass ones. This trend is consistent with self-similar scaling relations, and we find a mass dependence twice as strong as that suggested in Eq.~\eqref{eq:kSZ-scaling}. For a low-mass, relaxed sample of clusters, our rkSZ profile amplitudes are $\approx 30~\mu$K, consistent with \cite{2017MNRAS.465.2584B}.

	\item \textbf{Dynamical state.} Our metrics to assess the dynamical state of cluster atmospheres are correlated to the halo masses  (see Fig.~\ref{fig:slices:dynamical-state}) and indicate that unrelaxed clusters produce a rkSZ signal two times stronger than relaxed ones. While disentangling the dynamics of individual substructures in merging systems via the kSZ effect can be complex, stacking maps suppresses the effect of transient features and enhances the signal from bulk rotation. 

	\item \textbf{Spin alignment.} If stacking maps is a decisive step in retrieving the kSZ signal from cluster rotation, then de-rotating the maps and aligning the expected dipolar feature coherently are also crucial. This procedure relies on determining the spin orientation of the gas. We have shown that using the galaxy angular momentum orientation as a proxy for that of the hot gas, as implemented by \cite{2019JCAP...06..001B}, reduces the combined rkSZ signal by $\approx$ 60\% compared to when the de-rotation is based on the gas spin itself. This effect arises because galaxies and hot gas do not always co-rotate, as shown by a tail towards low values in the $\cos \theta_{04}$ distribution in Fig. \ref{fig:corner-plot-dynamical-state}. {Similarly, using the DM spin as a proxy for the gas spin would reduce the stacked rkSZ signal amplitude by about 40\%.}
	
	\item \textbf{Temperature power spectrum.} Adapting the formalism in \citetalias{CC02} for the temperature power spectrum calculation, we provide an improved method for estimating the one-halo term arising from cluster rotation. We remove the assumption of a generalised-NFW model and solid-body rotation and, instead, we input the rkSZ profiles from full-physics simulations. 
	%
	The description presented here can be directly used in, e.g., {\tt CLASS-SZ} \citep{Bolliet:2017lha} to compute the rkSZ temperature power spectrum contribution. While these are expected to be about one order of magnitude smaller than the usual kSZ effect \citepalias[see][]{CC02}, this could add a new cosmological noise-floor to studies of the moving lens effect \citep[e.g.,][]{2019PhRvL.123f1301H}. Future kSZ studies, e.g., with the Simons Observatory \citep{SOWP2019}, might also become sensitive to this additional kSZ component, with a possible contributions from internal substructure motion at small scales.
{Due to the difference in the amplitude of the rkSZ signal with respect to the model of \citetalias{CC02}, it will be important to consider these effects more carefully.}

\end{enumerate}
For microwave observations of the kSZ effect, an improved sensitivity may not guarantee a reliable reconstruction of the cluster rotation from the kSZ signal without a robust method of probing the orientation of the ICM spin. This step is critical for avoiding a potential $60$\% signal loss (see Section \ref{sec:results}) and observationally challenging \citep{2017MNRAS.465.2616M, 2019JCAP...06..001B}. {The \citet{2017MNRAS.465.2616M} sample consists of the most strongly rotating halos, as determined from galaxy LoS velocity data in SDSS-DR10. The resulting selection bias naturally favours strongly rotating halos, and could effectively lead to a signal loss smaller than we predicted without such selection. Provided that the spins of DM and galaxies are closely aligned, as we showed in Fig. \ref{fig:corner-plot-all-properties} for MACSIS, we can estimate the rkSZ signal suppression by using clusters with a strongly rotating DM halo (high $\lambda_{\rm DM}$) to represent those with a strongly rotating galaxy population. Indeed, the rkSZ amplitude of our high-$\lambda_{\rm DM}$ subset for the galaxies-aligned case is $59.5\, \mu$K, almost double that of the overall sample, $32.1\, \mu$K. On average, the signal suppression relative to $A_{\rm max}^{(\rm gas)}$ has a smaller impact on high-$\lambda_{\rm DM}$ clusters (42\%) compared to the overall sample (60\%, see Table~\ref{tab:max_measured}). Our study suggests that the object-selection strategy of \cite{2019JCAP...06..001B} can marginally boost the stacked rkSZ signal of galaxies-aligned clusters.}

Measuring the {LoS} velocity of the hot gas \textit{directly} may become feasible in the future with the advent of new high-resolution X-ray spectrometers \citep[see also][]{2013MNRAS.434.1565B}. The X-ray space observatory \textit{Athena}, developed by the European Space Agency \citep[see the white paper by][]{2013arXiv1306.2307N}, is planned to be launched in the late 2020s and will be equipped with the X-ray Integral Field Unit \citep[X-IFU,][]{2018SPIE10699E..1GB, 2018arXiv180706903G}. \textit{Athena}/X-IFU opens exciting prospects for a Doppler measurements of the {LoS} velocity field of the ICM, with a spectral resolution of $2.5-7$ eV in the soft X-ray band. These specifications are even superior to the capabilities of XRISM/Resolve micro-calorimeter array, managed by the Japan Aerospace Exploration Agency and capable of resolving Doppler speeds of $\simeq 300~{\rm km\, s^{-1}}$ \citep{2018arXiv180706903G, 2020SPIE11444E..22T, 2021JATIS...7c7001T} and, previously, of the Hitomi spectrometer which provided measurements of turbulent motions in the core of the Perseus cluster \citep{2016Natur.535..117H} by resolving speeds of $\simeq 100~{\rm km\, s^{-1}}$ at 6 keV \citep[][]{2018JATIS...4b1402T}.

More accurate predictions for the rkSZ amplitude will soon be made possible by future large-volume hydrodynamic simulations. In the very near future, an example of such runs is Virgo Consortium's flagship FLAMINGO project, which will contain a much richer cluster sample of $\sim 10^6$ objects modelled with full physics \citep{flamingo_schaye2023}. With clusters as massive as the MACSIS objects and a mass-limited HMF down to $M_{500}\simeq 10^{13}$ M$_\odot$, FLAMINGO will be able to reproduce cluster-count statistics and improve the estimates of our rkSZ profiles using particle-data snapshots and halo catalogues. In addition, the light-cone outputs from FLAMINGO will allow to construct Healpix \citep{2005ApJ...622..759G} all-sky maps over a selected redshift range. In a future work, we aim to use these data sets to forecast the rotational kSZ signal from clusters through power spectra and feature extraction methods \citep[see e.g.][]{1996MNRAS.279..545H, 2021MNRAS.507.4852Z}. {Together with an improved modelling of the relativistic SZ effect \citep[e.g.][]{2022MNRAS.517.5303L} this could refine the simulation-driven prescription of SZ clusters in cosmology.}

As an observational outlook, we will also illustrate how our framework could facilitate the study of CMB foregrounds and play a role in upcoming precision cosmology programs with stage-4 facilities, such as Simons Observatory \citep{SOWP2019} and SKA-2 (see \citealt{2016PhRvD..94d3522A} for a review, and \citealt{2016arXiv161002743A}). In particular, estimates from simulations could guide the development of four additional areas of research: matched filters for separating the cluster rotation from the moving-lens effect \citep{1986Natur.324..349G, 2007MNRAS.380.1023S, 2021PhRvD.104h3529H, 2021PhRvD.103d3536H}; the pairwise transverse velocity measurement with the Rees-Sciama effect \citep{2019ApJ...873L..23Y, 2019PhRvL.123f1301H}; the vector gravito-magnetic distortion, predicted to occur when rotating massive clusters induce space-time frame-dragging \citep{2021ApJ...911...44T, 2022MNRAS.510.3589B}; and the $\simeq 10\, \sigma$ measurement of cosmic filament rotation using the kSZ effect \citep{2023MNRAS.519.1171Z}.


\section*{Acknowledgements}

We thank Xuelei Chen and Martin Murin for insightful discussions. This work used the DiRAC@Durham facility managed by the Institute for Computational Cosmology on behalf of the STFC DiRAC HPC Facility (\href{www.dirac.ac.uk}{www.dirac.ac.uk}). The equipment was funded by BEIS capital funding via STFC capital grants ST/K00042X/1, ST/P002293/1, ST/R002371/1 and ST/S002502/1, Durham University and STFC operations grant ST/R000832/1. DiRAC is part of the National e-Infrastructure. EA and IT acknowledge the STFC studentship grant ST/T506291/1. 
{JC was furthermore supported by the ERC Consolidator Grant {\it CMBSPEC} (No.~725456) and the Royal Society as a Royal Society University Research Fellow at the University of Manchester, UK (No.~URF/R/191023).}
The research in this paper made use of the following software packages and libraries: 
\textsc{Python} \citep{van1995python},
\textsc{Numpy} \citep{harris2020array},
\textsc{Scipy} \citep{virtanen2020scipy},
\textsc{Numba} \citep{lam2015numba},
\textsc{Matplotlib} \citep{hunter2007matplotlib, caswell2020matplotlib},
\textsc{SWIFTsimIO} \citep{Borrow2020} and
\textsc{Astropy} \citep{robitaille2013astropy, price2022astropy},
\textsc{Unyt} \citep{goldbaum2018unyt}. The colour scheme used throughout the document was generated using the open-source tool \texttt{ColorBrewer} \citep{harrower2003colorbrewer}.

\section*{Data Availability}
The MACSIS simulations were produced by \cite{macsis_barnes_2017}. Enquiries concerning the availability of raw snapshot data, \texttt{SUBFIND} halo catalogues and the version of the \texttt{Gadget-3} code used to run the simulations should be directed to the original authors. The code used for reading and querying the MACSIS data is publicly available on the first author's GitHub repository (\href{https://github.com/edoaltamura/macsis-cosmosim}{github.com/edoaltamura/macsis-cosmosim}) and we include the data products used to generate the figures presented throughout the document. The repository also contains the list of indices of the clusters in the $z=0$ and $z=1$ subsets of Section \ref{sec:results:z-dependence}, ordered by the  FoF of the MACSIS parent simulation. Additional information for reproducing our results is contained in \texttt{JSON} files; e.g. the parameters for the density profile model and the median values of $\{r_{500},\, r_{200},\, M_{500},\, M_{200},\, \lambda_{\rm gas},\, v_{\rm circ}\}$ for each selection subset. The bootstrap samples are not included, but can be reproduced using the code provided. Some intermediate data products, such as original rkSZ cluster maps, are too large to be hosted on GitHub and can be made available upon request to the corresponding authors.

\bibliographystyle{mnras}
\bibliography{main} 


\appendix

\section{Correlations of cluster properties}
\label{app:correlation-coefficients}

\subsection{Analytic relation between \texorpdfstring{$\alpha_{\rm 3D}$}{alpha} and \texorpdfstring{$\beta_{\rm 3D}$}{beta}}
\label{app:correlation:alpha-beta}
To justify the correlation in Fig. \ref{fig:corner-plot-dynamical-state} between the non-thermal pressure fraction, $\alpha$, and the kinetic-to-thermal ratio, $\beta$, we prove the relation between these quantities introduced in Eq. \eqref{eq:alpha}.
We summarise the definitions of the kinetic and thermal energy:
\begin{subequations}
\begin{align}
    E_{\rm kin} &= \frac{1}{2} \sum_i m_i ({\bf v}_i - {\bf v_{\rm bulk}})^2
    \label{eq:energies-summary:kinetic}
    \\[1mm]
    E_{\rm th} &= \frac{3}{2}\,{\rm k_B}\sum_i \frac{T_i\, m_i}{\mu\,m_{\rm P}}.
    \label{eq:energies-summary:thermal}
\end{align}
\end{subequations}
We begin by studying the form of the equation for the kinetic energy. Since we compute the bulk velocity in the set of particles $\{i\}$, we have the expression
\begin{equation}
    \label{eq:bulk-motion-app}
    {\bf v_{\rm bulk}} = \frac{\sum_i m_i {\bf v}_i}{\sum_i m_i} = \frac{\sum_i m_i {\bf v}_i}{M} \equiv \langle {\bf v} \rangle = \sum_j \langle v_j \rangle,
\end{equation}
where $M\equiv \sum_i m_i$. By substituting the above in Eq. \eqref{eq:energies-summary:kinetic}, and expanding the calculation for each spatial component $j\in\{x, y, z\}$, one can obtain
\begin{subequations}
\begin{align}
    E_{{\rm kin}, j} &= \frac{1}{2} \sum_i m_i \left(  v_{i,j} - \langle v_j \rangle \right)^2
    \\[1mm]
    &= \frac{1}{2} \sum_i m_i \left(v_{i,j}^2 - 2\, v_{i,j} \langle v_j \rangle + \langle v_j \rangle ^ 2 \right)
    \\[1mm]
    &= \frac{1}{2} \biggl[ 
    \sum_i m_i v_{i,j}^2 - 2\, \langle v_j \rangle \underbrace{\sum_i m_i v_{i,j}}_{\substack{\text{$=M\,\langle v_j \rangle$} \\ \text{from Eq.~\eqref{eq:bulk-motion-app}}}}  +~ \langle v_j \rangle ^ 2 \underbrace{\sum_i m_i}_{\text{$=M$}} \biggr]
    \\[2mm]
    &= \frac{1}{2} \biggl[ \sum_i m_i v_{i,j}^2 - 2\, M \langle v_j \rangle^2 + M \langle v_j \rangle ^ 2 \biggr]
    \\[1mm]
    &= \frac{1}{2} \left( \sum_i m_i v_{i,j}^2 - M \langle v_j \rangle ^ 2 \right)
    \label{eq:kinetic-calculation-last}
    \\[1mm]
    &= \frac{1}{2} M\left(\langle v_{j}^2 \rangle - \langle v_j \rangle ^ 2 \right).
    \label{eq:kinetic-calculation-final}
\end{align}
\end{subequations}
The first term in Eq. \eqref{eq:kinetic-calculation-last} is just the mass-weighted square of the velocities in $\{i\}$ for the component $j$, which can be written as $\sum_i m_i v_{i,j}^2 = M \langle v_{j}^2 \rangle$, in analogy to Eq. \eqref{eq:bulk-motion-app}.
Eq. \eqref{eq:kinetic-calculation-final} contains the definition of the velocity dispersion $\sigma_j^2 \equiv \langle v_{j}^2 \rangle - \langle v_j \rangle ^ 2$. Finally, we can sum the kinetic energy for the three components of the velocity dispersion:
\begin{equation}
    \label{eq:kinetic-vel-dispersion}
    E_{\rm kin} = \frac{1}{2} M \sum_j \sigma_j^2 = \frac{1}{2} M (\sigma_x^2 + \sigma_y^2 + \sigma_z^2) \equiv \frac{1}{2} M \sigma^2.
\end{equation}

The next step involves rearranging Eq. \eqref{eq:energies-summary:thermal} and expressing the thermal energy in terms of the mass-weighted temperature for the ensemble of particles $\{i\}$
\begin{equation}
    \label{eq:mass-weighted-temperature}
    T_{\rm mw} = \frac{\sum_i m_i\, T_i}{\sum_i m_i} = \frac{\sum_i m_i\, T_i}{M}.
\end{equation}
This procedure is carried out as follows:
\begin{subequations}
\begin{align}
    E_{\rm th} &= \frac{3}{2}\,{\rm k_B}\sum_i \frac{T_i\, m_i}{\mu\,m_{\rm P}}
    \\[1mm]
    &= \frac{3}{2}\,\frac{\rm k_B}{\mu\,m_{\rm P}}\, \sum_i T_i\, m_i
    \\[1mm]
    &= \frac{3}{2}\,\frac{{\rm k_B}\, T_{\rm mw}}{\mu\,m_{\rm P}}\,M.
    \label{eq:thermal-energy-derivation}
\end{align}
\end{subequations}
Combining Eqs. \eqref{eq:kinetic-vel-dispersion} and \eqref{eq:thermal-energy-derivation} gives
\begin{equation}
    \beta = \frac{E_{\rm kin}}{E_{\rm th}} = \frac{\mu\,m_{\rm P}\, \sigma^2}{3\, {\rm k_B}\, T_{\rm mw}}.
\end{equation}
After defining the non-thermal pressure as 
\begin{equation}
    P_{\rm nth} = \frac{1}{3}\, \rho_{\rm gas} \sigma^2,
\end{equation}
where $\rho_{\rm gas}$ the local density of the hot gas \citep[e.g.][]{2014MNRAS.442..521S, 2022arXiv221101239T}, we can reproduce the formulation in Eq. \eqref{eq:alpha}
\begin{subequations}
\begin{align}
    \alpha &= \frac{P_{\rm nth}}{P_{\rm nth} + P_{\rm th}} 
    \\[1mm]
    &= \frac{1/3\, \rho \sigma^2}{1/3\, \rho \sigma^2 + \rho\, {\rm k_B}\, T_{\rm mw}/(\mu\,m_{\rm P})}
    \label{eq:alpha-derivation-gas-density}
    \\[2mm]
    &= \biggl[ 1 + \underbrace{\frac{3\, {\rm k_B}\, T_{\rm mw}}{\mu\,m_{\rm P}\, \sigma^2}}_{=1/\beta} \biggr]^{-1},
\end{align}
\end{subequations}
where $P_{\rm th}$ is the thermal pressure, derived from the ideal gas law. Rearranging the expression above, we finally obtain
\begin{equation}
    \alpha = \frac{\beta}{1 + \beta}.
\end{equation}
This relationship can explain the large correlation between the $\alpha$ quantities and $\beta_{\rm 3D}$ reported in Fig. \ref{fig:corner-plot-dynamical-state} and Appendix \ref{app:correlation-coefficients}. Furthermore, we note that $\beta_{\rm 3D}$ is a cluster-averaged property, since it is obtained for all hot gas particles in $r_{500}$, while $\alpha_{\rm 3D}(r)$ is a profile evaluated from particles in a thin spherical shell at radius $r$.

\subsection{Correlation coefficients}

\begin{figure*}
    \centering
	\includegraphics[width=2\columnwidth]{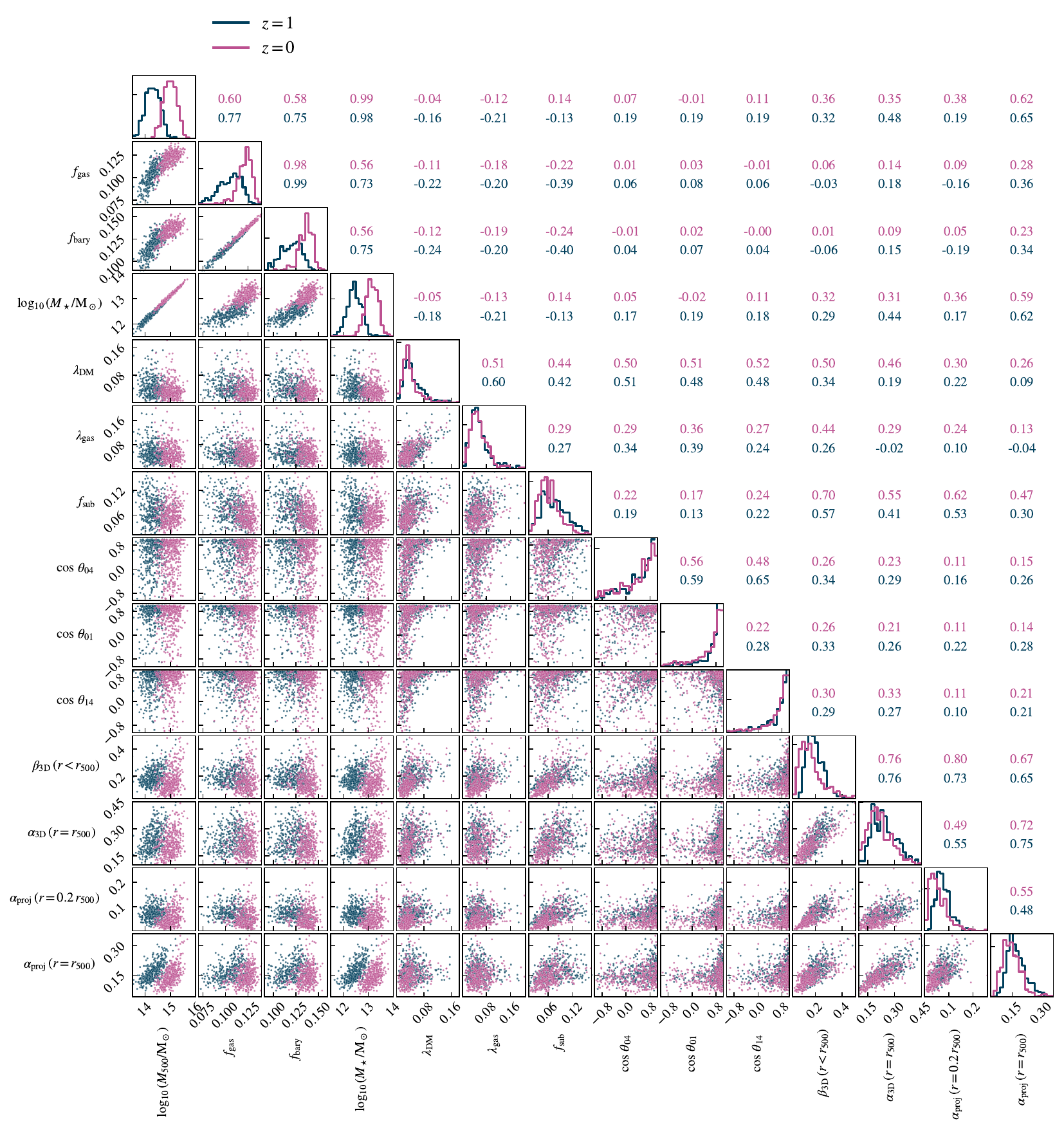}
    \caption{As in Fig. \ref{fig:corner-plot-basic-properties}. In the top right, we write the Spearman rank correlation coefficients for each data set pair, colour-coded by redshift.}
    \label{fig:corner-plot-all-properties}
\end{figure*}

In addition to the corner plots in Figs. \ref{fig:corner-plot-basic-properties} and \ref{fig:corner-plot-dynamical-state}, we provide the complete set of correlation relations in Fig. \ref{fig:corner-plot-all-properties}. Here, we also quote the Spearman correlation coefficient, a parameter that quantifies the correlation between any two given sets of data, assuming that a monotonic relation is expected. The Spearman correlation coefficient of two variables $X$ and $Y$ is defined by
\begin{equation}
    r_{\rm S} = 1- {\frac {6 \sum d_i^2(X,Y)}{n(n^2 - 1)}},
\end{equation}
where $n$ is the number of data points, or realisations, and $d_i(X,Y)$ is the pairwise distances of the ranks of the variables $X_i$ and $Y_i$.

In Fig. \ref{fig:corner-plot-all-properties}, we also report the distributions for $\cos\, \theta_{01}$, corresponding to the angle between gas and dark matter components, and $\cos\, \theta_{14}$ for the angle between dark matter and stars. We can recover high values of $r_{\rm S} \simeq 0.6-1$ for known mass-scaling relations, such as gas fraction, baryon fraction and stellar mass in $r_{500}$. We find small, negative correlations for $\lambda_{\rm DM}$ and  $\lambda_{\rm gas}$ against $M_{500}$, in agreement with \cite{2010MNRAS.404.1137B}. The correlation coefficient between $\alpha_{\rm 3D}$ (at a particular radius) and $\beta_{\rm 3D}$ (integrated over $r_{500}$) quantities is $r_{\rm S} \simeq 0.6-0.7$, as expected from Appendix \ref{app:correlation:alpha-beta}. We also report significantly high correlations between $f_{\rm sub}$ and the quantities based on $\alpha_{\rm 3D}$ and $\beta_{\rm 3D}$, suggesting that MACSIS clusters with high substructure fractions tend to have a thermodynamically unrelaxed and turbulent atmosphere.

\section{Analytic fits to the rkSZ profiles}
\label{app:profile-fits}
\begin{figure}
    \includegraphics[width=\columnwidth]{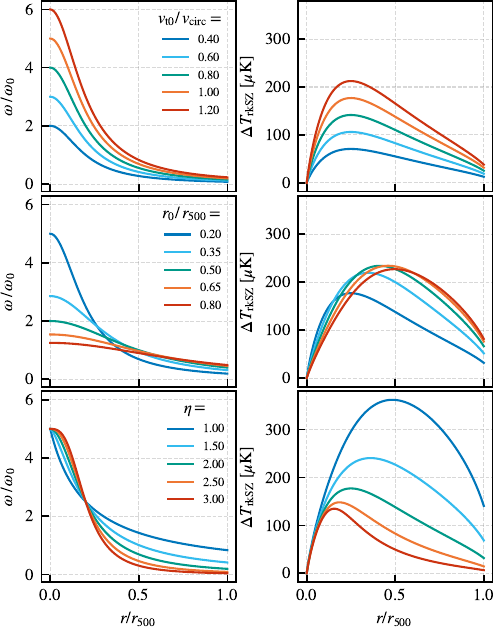}
    \caption{Effect of varying the parameter values in Eq.~\eqref{eq:omega-profile} on the angular velocity profile (left) and the rkSZ profile (right), assuming a \protect\cite{2006ApJ...640..691V} electron number density profile from the MACSIS 0 cluster at $z=0$. From top to bottom, we vary the amplitude $v_{\rm t0}$, the scale-radius $r_0$ and the slope $\eta$. The model with $\eta = 2$ corresponds to the \protect\cite{2017MNRAS.465.2584B} functional form.}
    \label{fig:generalised-baldi-model}
\end{figure}

To match the simulated rkSZ profiles with a analytic functional form, we adopt angular velocity profile which depends on three free parameters, as in Eq.~\eqref{eq:omega-profile}. We found that the $\omega(r)$ model in \cite{2017MNRAS.465.2584B} could not fit the slope of the MACSIS rkSZ profiles beyond $r_{\rm max}$. Therefore, we introduced the $\eta$ parameter, which controls the slope of the $\omega(r)$ profile with a pivot at $r_0$. In Fig.~\ref{fig:generalised-baldi-model}, we illustrate how $\omega(r)$ changes when varying each parameter individually, and the effect on the rkSZ profiles. For convenience, we normalise the $\omega(r)$ profile by the constant $\omega_0 \equiv v_{\rm circ}/ r_{500} = \sqrt{GM_{500}/r_{500}^3}$. For each column, we set $\mathcal{B}_{\rm g}=\{v_{\rm t0} = v_{\rm circ},\, r_0=r_{500}/5,\, \eta = 2 \}$ as default parameters and we only allow one parameter to vary. In the top panels, we show the variation of the tangential velocity scale, $v_{\rm t0}$, which controls the overall amplitude of the profiles. In the middle panels, we show that increasing the scale radius $r_0$ causes the $\omega(r)$ to become shallower, alters the maximum rkSZ amplitude $A_{\rm max}$ and changes the profile of the slope beyond $r_{\rm max}$. In the bottom panels, we show that the rkSZ amplitude decays to 0 $\mu$K faster with radius for higher values of $\eta$ ($\gtrsim 2$). By setting $\eta = 2$, we recover the model used in \cite{2017MNRAS.465.2584B}:
\begin{equation}
    \omega(r) = \frac{v_{\rm t0}}{r_0 \left[1 + (r/r_0)^2\right]}.
\end{equation}
For {the limiting case where} $\eta \longrightarrow 1$, the angular velocity profile becomes cusp-like at $r\simeq 0$; the rate of decay $d\omega/dr$ drops rapidly and $\omega(r)$ is still relatively large at $r \simeq r_{500}$. Finally, setting $\eta = 0$ yields $\omega(r) = v_{\rm t0} / (2\, r_0) \sim {\rm constant}$, similarly to the solid-body rotation model assumed by \citetalias{CC02} and \citetalias{CM02}.

\begin{figure}
    \includegraphics[width=\columnwidth]{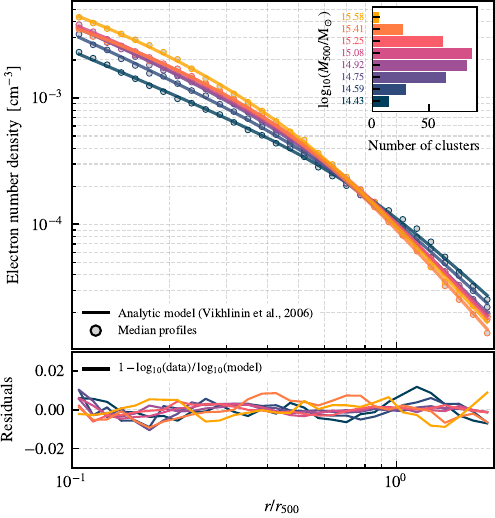}
    \caption{\textit{Top.} Median density profiles in 8 log-spaced mass bins and their analytic fits (solid lines) using a \protect\cite{2006ApJ...640..691V} model with 5 free parameters ($\alpha$ is fixed to 1.5). The inset at the top-right is the HMF of the MACSIS $z=0$ sample (as in Fig. \ref{fig:slices:properties}.A), shown using eight $M_{500}$ logarithmic bins. The central value of the bins is shown on the $y$-axis and the number of clusters in each bin on the $x$-axis. The bins are colour coded to match the fits in the main panel and the data for the median profiles. {\textit{Bottom.} Logarithmic percentile residuals taking the analytic model as baseline. The fits are well converged and the measured median profiles never deviate more than 1\% from the best fit model.}}
    \label{fig:density-profile-mass-bins}
\end{figure}

\begin{figure}
    \includegraphics[width=\columnwidth]{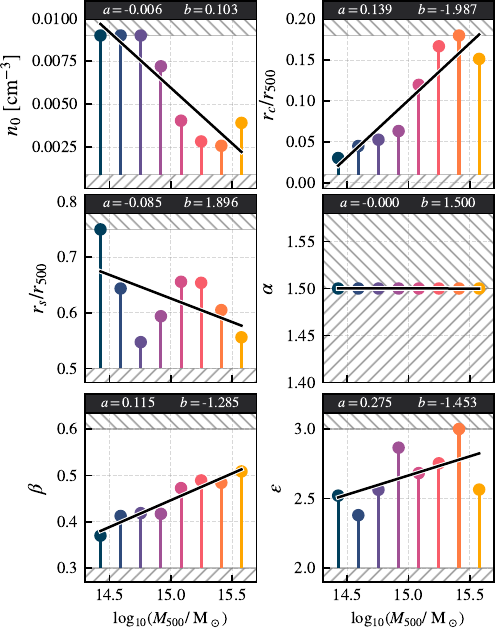}
    \caption{Summary of the $\mathcal{V}_6$ best-fit parameters for the electron density profiles shown in Fig. \ref{fig:density-profile-mass-bins} with increasing halo mass. The data points are coloured by mass bin as in Fig. \ref{fig:density-profile-mass-bins} and we indicate the upper and lower bounds imposed to the fit with hatched regions. For $\alpha$, the bounds edges coincide, allowing only for $\alpha=1.5$. For each parameter, we fit a linear model (black solid line) and we report the gradient $a$ and the intercept $b$ at the top of the corresponding panel.}
    \label{fig:density-profile-mass-bins-parameters}
\end{figure}

To fit the rkSZ profile template to the simulation data, we use the best-fit \cite{2006ApJ...640..691V} parameters, $\mathcal{V}_6$, as priors. We report their values in Table~\ref{tab:vikhlinin_z0} for each subset of the MACSIS sample. In addition to the selection methods used throughout Section \ref{sec:results}, we split the cluster population in 8 logarithmic mass bins and fit the median density profile for each bin, as shown in Fig.~\ref{fig:density-profile-mass-bins}. In the left panel, we show the median density profiles (circles) for each mass bin, fit by a  \cite{2006ApJ...640..691V} model. On the top-right, the HMF shows that most of the mass bins have a limited number of objects ($\sim 10$). To avoid over-fitting to transient features of the density profile, we reduced the dimensionality of the original \cite{2006ApJ...640..691V} parameter space to 5 free parameters, and we set $\alpha=1.5$. The values of the parameters $\mathcal{V}_6$ for each mass bin, shown on the right of Fig.~\ref{fig:density-profile-mass-bins-parameters}, provided the simple, linear mass-scaling relations reported in Eq.~\eqref{eq:Ne_approximation} with gradient and slope values listed in Table~\ref{tab:mass-dependence-fit-params}. In this work, these scaling relations are only used to provide guidelines for the temperature power spectrum calculation (see Section \ref{sec:power-spectrum}).

Given the $\mathcal{V}_6$ parameters in Table~\ref{tab:vikhlinin_z0}, we report the best-fit parameters $\mathcal{B}_{\rm g}$ for the analytic rkSZ profile in Table~\ref{tab:baldi_z0}, based on the method in Section \ref{sec:results:fitting}. {In addition to the value of the tangential scale velocity, we provide a parametrisation similar to the $\omega$ profile in Eq.~(10) of \citetalias{CC02}: $v_{\rm t0} = 3\, \xi_0 \lambda_{\rm gas} v_{\rm circ}$, where we express the spin parameter scaling explicitly and we introduce a positive coefficient $\xi_0$ as a free parameter.}
We commonly find $\xi_0\simeq 10$, which demonstrates that the simulation clusters rotate faster than assumed in \citetalias{CC02}. To give a reference, by comparing to $\omega_0=v_{\rm circ}/r_{500}$ (chosen in Fig.~\ref{fig:generalised-baldi-model}) for $\eta=2$ and $r_0=r_{500}/5$ we expect $\omega(r=0)/\omega_0= v_{\rm t0}/(r_0 \omega_0) = 5\, v_{\rm t0}/v_{\rm circ}$ as found in the upper left panel of Fig.~\ref{fig:generalised-baldi-model}. From the fits in Table~\ref{tab:baldi_z0}, we furthermore conclude that $3\, \xi_0 \lambda_{\rm gas}  \simeq 1.5$ for the gas-aligned case, which is about ten times larger than \citetalias{CC02}.


\begin{table}
    \setlength{\tabcolsep}{3pt}
    \centering
    \caption{Summary of the selection criteria (column 1) and the best fit parameters from the median electron number density profile using the \protect\cite{2006ApJ...640..691V} model (columns 2-6), with fixed $\alpha=1.5$. These results are obtained at $z=0$ unless stated otherwise.}
    \label{tab:vikhlinin_z0}
    \rowcolors{5}{orange!20}{white}
    \begin{tabular}{lccccc}
    \toprule
    \multicolumn{1}{c}{}    &   
    \multicolumn{5}{c}{\bf Vikhlinin parameters} \\ 
    \cline{2-6} \rule{0pt}{1ex}
    
    Selection criterion     &  $n_0$ &  $r_c$ &  $r_s$ &  $\beta$ &  $\varepsilon$ \\
                            &  [$10^{-3}~{\rm cm}^{-3}$] & [$r_{500}$] & [$r_{500}$] & [--] & [--] \\
    \midrule
    \rowcolor{gray!25}
All clusters                                       &    4.49 &   0.10 &      0.61 &     0.44 &   2.74\\
$M_{500} < 9.7\times 10^{14}$ M$_\odot$            &    9.00 &   0.05 &      0.60 &     0.42 &   2.63\\
$M_{500} > 9.7\times 10^{14}$ M$_\odot$            &    2.70 &   0.18 &      0.70 &     0.52 &   2.61\\
$f_{\rm gas}$ < 0.12                               &    2.69 &   0.15 &      0.75 &     0.48 &   2.49\\
$f_{\rm gas}$ > 0.12                               &    8.59 &   0.07 &      0.64 &     0.45 &   3.00\\
$f_{\rm bary}$ < 0.14                              &    4.50 &   0.09 &      0.62 &     0.41 &   2.68\\
$f_{\rm bary}$ > 0.14                              &    5.70 &   0.09 &      0.61 &     0.46 &   2.80\\
$M_{\star} < 1.4\times 10^{13}$ M$_\odot$          &    9.00 &   0.05 &      0.59 &     0.42 &   2.61\\
$M_{\star} > 1.4\times 10^{13}$ M$_\odot$          &    2.62 &   0.18 &      0.68 &     0.51 &   2.66\\
$\lambda_{\rm DM}$ < 0.03                          &    9.00 &   0.05 &      0.50 &     0.41 &   2.63\\
$\lambda_{\rm DM}$ > 0.03                          &    3.16 &   0.13 &      0.73 &     0.46 &   3.00\\
$\lambda_{\rm gas}$ < 0.05                         &    7.97 &   0.07 &      0.53 &     0.43 &   2.68\\
$\lambda_{\rm gas}$ > 0.05                         &    3.09 &   0.13 &      0.72 &     0.45 &   2.85\\
    \midrule
$\beta_{\rm 3D}$ < 0.15                            &    9.00 &   0.06 &      0.52 &     0.44 &   2.36\\
$\beta_{\rm 3D}$ > 0.15                            &    2.22 &   0.18 &      0.73 &     0.47 &   3.00\\
$f_{\rm sub}$ < 0.06                               &    9.00 &   0.06 &      0.52 &     0.45 &   2.39\\
$f_{\rm sub}$ > 0.06                               &    2.30 &   0.17 &      0.75 &     0.47 &   3.00\\
$\cos\, \theta_{04}$ < 0.56                        &    4.95 &   0.09 &      0.60 &     0.44 &   2.60\\
$\cos\, \theta_{04}$ > 0.56                        &    4.09 &   0.10 &      0.61 &     0.43 &   2.87\\
$\alpha_{\rm 3D}$ < 0.21                           &    4.80 &   0.10 &      0.59 &     0.46 &   2.45\\
$\alpha_{\rm 3D}$ > 0.21                           &    3.36 &   0.13 &      0.69 &     0.45 &   2.99\\
$\alpha_{\rm proj}\,(r= 0.2\,r_{500})$ < 0.05      &    9.00 &   0.06 &      0.54 &     0.46 &   2.34\\
$\alpha_{\rm proj}\,(r= 0.2\,r_{500})$ > 0.05      &    2.13 &   0.18 &      0.69 &     0.45 &   3.00\\
$\alpha_{\rm proj}\,(r=r_{500})$ < 0.14            &    9.00 &   0.05 &      0.56 &     0.42 &   2.53\\
$\alpha_{\rm proj}\,(r=r_{500})$ > 0.14            &    2.51 &   0.18 &      0.71 &     0.50 &   2.77\\
    \midrule
$z=0$ (75 clusters)                                &    9.00 &   0.05 &      0.61 &     0.41 &   2.42\\
$z=1$ (70 clusters)                                &    4.59 &   0.28 &      0.70 &     0.53 &   3.00\\
    \bottomrule
    \end{tabular}
\end{table}

\begin{table*}
\setlength{\tabcolsep}{4pt}
    \centering
    \caption{{Summary of the halo parameters and best fit parameters for the analytic rkSZ profiles. From left to right: selection criteria defining the cluster sub-sample (first column); the median value of the hot gas spin parameter, $\lambda_{\rm gas}$, and the circular velocity, $v_{\rm circ}$, at $r_{500}$ computed from the simulations for each sub-sample (\textit{subset medians}); the best fit parameters for the $\omega(r)$ profile from the \textit{gas}-edge-on projection, the \textit{galaxies}-edge-on projection and the \textit{DM}-edge-on projection. These results are obtained at $z=0$ unless stated otherwise.}}
    \label{tab:baldi_z0}
    \rowcolors{9}{orange!20}{white}
    \begin{tabular}{lccc|cccc|cccc|cccc}
    \toprule
    
    \multicolumn{1}{c}{}    &
    \multicolumn{2}{c}{\bf Subset medians}    & 
    \multicolumn{4}{c}{\bf Gas-aligned} & 
    \multicolumn{4}{c}{\bf Galaxies-aligned} & 
    \multicolumn{4}{c}{\bf DM-aligned}\\ 
    
    \cmidrule(rl){2-3} \cmidrule(rl){4-7} \cmidrule(rl){8-11} \cmidrule(rl){12-15} \rule{0pt}{1ex}    
    
     Selection criterion   & $\lambda_{\rm gas}$ & $v_{\rm circ}$ &  $v_{\rm t0}$ & $\xi_0$ & $r_0$ & $\eta$  & $v_{\rm t0}$ & $\xi_0$ &  $r_0$ & $\eta$ &  $v_{\rm t0}$ & $\xi_0$ & $r_0$ & $\eta$ \\ 
                           &  [-] & [km/s] & [$v_{\rm circ}$]   &  [-]  & [$r_{500}$]   & [-]  & [$v_{\rm circ}$] & [-]  & [$r_{500}$]   & [-]  & [$v_{\rm circ}$] & [-]  & [$r_{500}$]  & [-]  \\
    \midrule
    \rowcolor{gray!25}
All clusters                                  & 0.0505 &   1648 & 1.53 &    10.02 & 0.16 & 2.01 & 0.51 & 3.35 & 0.16 & 1.82 &  0.94 & 6.19 & 0.11 & 1.80 \\
$M_{500} < 9.7\times 10^{14}$ M$_\odot$       & 0.0542 &   1430 & 0.99 &    6.06  & 0.17 & 2.25 & 0.30 & 1.84 & 0.18 & 2.00 &  0.71 & 4.36 & 0.12 & 2.00 \\
$M_{500} > 9.7\times 10^{14}$ M$_\odot$       & 0.0462 &   1891 & 1.92 &    13.85 & 0.15 & 1.92 & 0.68 & 4.88 & 0.14 & 1.75 &  1.14 & 8.19 & 0.10 & 1.70 \\
$f_{\rm gas}$ < 0.12                          & 0.0532 &   1477 & 0.93 &    5.84  & 0.15 & 1.70 & 0.27 & 1.70 & 0.53 & 2.34 &  0.41 & 2.55 & 0.25 & 1.50 \\
$f_{\rm gas}$ > 0.12                          & 0.0462 &   1807 & 1.47 &    10.58 & 0.12 & 1.92 & 0.61 & 4.41 & 0.10 & 1.81 &  1.16 & 8.37 & 0.09 & 1.87 \\
$f_{\rm bary}$ < 0.14                         & 0.0547 &   1493 & 1.46 &    8.88  & 0.21 & 2.15 & 0.46 & 2.82 & 0.40 & 2.46 &  0.71 & 4.31 & 0.24 & 2.05 \\
$f_{\rm bary}$ > 0.14                         & 0.0458 &   1779 & 1.53 &    11.12 & 0.13 & 2.01 & 0.64 & 4.62 & 0.09 & 1.83 &  1.17 & 8.54 & 0.08 & 1.84 \\
$M_{\star} < 1.4\times 10^{13}$ M$_\odot$     & 0.0522 &   1430 & 0.55 &    3.51  & 0.09 & 1.69 & 0.16 & 1.03 & 0.07 & 1.48 &  0.43 & 2.76 & 0.05 & 1.60 \\
$M_{\star} > 1.4\times 10^{13}$ M$_\odot$     & 0.0474 &   1891 & 2.10 &    14.74 & 0.16 & 1.96 & 0.74 & 5.18 & 0.15 & 1.74 &  1.25 & 8.78 & 0.10 & 1.72 \\
$\lambda_{\rm DM}$ < 0.03                     & 0.0433 &   1649 & 0.69 &    5.31  & 0.11 & 1.61 & 0.16 & 1.24 & 0.05 & 1.55 &  0.36 & 2.79 & 0.05 & 1.56 \\
$\lambda_{\rm DM}$ > 0.03                     & 0.0619 &   1644 & 2.49 &    13.39 & 0.16 & 2.13 & 1.31 & 7.03 & 0.13 & 1.83 &  1.73 & 9.30 & 0.14 & 1.93 \\
$\lambda_{\rm gas}$ < 0.05                    & 0.0348 &   1702 & 0.70 &    6.67  & 0.23 & 2.23 & 0.18 & 1.68 & 0.29 & 2.32 &  0.39 & 3.76 & 0.05 & 1.58 \\
$\lambda_{\rm gas}$ > 0.05                    & 0.0725 &   1596 & 2.50 &    11.48 & 0.12 & 1.85 & 1.13 & 5.18 & 0.09 & 1.62 &  1.68 & 7.70 & 0.13 & 1.85 \\
\midrule
$\beta_{\rm 3D}$ < 0.15                       & 0.0417 &   1562 & 0.80 &    6.43  & 0.10 & 1.92 & 0.23 & 1.87 & 0.27 & 2.70 &  0.40 & 3.17 & 0.22 & 2.44 \\
$\beta_{\rm 3D}$ > 0.15                       & 0.0584 &   1761 & 2.66 &    15.18 & 0.21 & 2.20 & 0.91 & 5.21 & 0.08 & 1.53 &  1.78 & 10.16 & 0.08 & 1.68 \\
$f_{\rm sub}$ < 0.06                          & 0.0450 &   1638 & 0.91 &    6.70  & 0.12 & 2.08 & 0.28 & 2.05 & 0.17 & 2.27 &  0.59 & 4.36 & 0.18 & 2.55 \\
$f_{\rm sub}$ > 0.06                          & 0.0556 &   1703 & 2.60 &    15.56 & 0.24 & 2.41 & 1.03 & 6.16 & 0.18 & 1.94 &  1.47 & 8.79 & 0.08 & 1.63 \\
$\cos\, \theta_{04}$ < 0.56                   & 0.0439 &   1633 & 1.22 &    9.23  & 0.11 & 1.90 & 0.37 & 2.78 & 0.12 & 2.34 &  0.47 & 3.55 & 0.05 & 1.52 \\
$\cos\, \theta_{04}$ > 0.56                   & 0.0581 &   1681 & 2.04 &    11.70 & 0.18 & 2.09 & 1.62 & 9.27 & 0.18 & 2.08 &  1.63 & 9.35 & 0.14 & 1.97 \\
$\alpha_{\rm 3D}$ < 0.21                      & 0.0441 &   1559 & 0.99 &    6.06  & 0.17 & 2.25 & 0.30 & 1.85 & 0.18 & 2.01 &  0.71 & 4.36 & 0.11 & 2.00 \\
$\alpha_{\rm 3D}$ > 0.21                      & 0.0578 &   1759 & 1.92 &    13.85 & 0.15 & 1.92 & 0.68 & 4.88 & 0.14 & 1.75 &  1.14 & 8.19 & 0.10 & 1.70 \\
$\alpha_{\rm proj}\,(r= 0.2\,r_{500})$ < 0.05 & 0.0461 &   1536 & 0.93 &    5.84  & 0.15 & 1.70 & 0.27 & 1.69 & 0.53 & 2.34 &  0.41 & 2.55 & 0.25 & 1.51 \\
$\alpha_{\rm proj}\,(r= 0.2\,r_{500})$ > 0.05 & 0.0541 &   1778 & 1.47 &    10.58 & 0.12 & 1.92 & 0.61 & 4.41 & 0.10 & 1.82 &  1.16 & 8.36 & 0.09 & 1.87 \\
$\alpha_{\rm proj}\,(r=r_{500})$ < 0.14       & 0.0480 &   1484 & 1.46 &    8.88  & 0.21 & 2.15 & 0.46 & 2.82 & 0.40 & 2.46 &  0.71 & 4.31 & 0.24 & 2.10 \\
$\alpha_{\rm proj}\,(r=r_{500})$ > 0.14       & 0.0531 &   1855 & 1.53 &    11.12 & 0.13 & 2.00 & 0.64 & 4.62 & 0.09 & 1.83 &  1.17 & 8.54 & 0.08 & 1.84 \\
\midrule
$z=0$ (matched HMF, 75 clusters)              & 0.0560 &   1277 & 0.98 &    6.05  & 0.17 & 2.24 & 0.30 & 1.84 & 0.18 & 2.00 &  0.71 & 4.36 & 0.11 & 2.00 \\
$z=1$ (matched HMF, 70 clusters)              & 0.0430 &   1567 & 1.92 &    13.85 & 0.14 & 1.92 & 0.68 & 4.88 & 0.14 & 1.75 &  1.14 & 8.19 & 0.10 & 1.70 \\
    \bottomrule
    \end{tabular}
\end{table*}

\bsp	
\label{lastpage}
\end{document}